\documentclass[aps,prb,twocolumn,superscriptaddress]{revtex4-1}

\usepackage[usenames,dvipsnames]{xcolor}
\usepackage[caption=false]{subfig} 
\usepackage{pgf,tikz}  

\graphicspath{{figures/}}

\usepackage{csquotes} 
\usepackage{upgreek} 
\usepackage[version=3]{mhchem} 

\usepackage[pdftex,colorlinks,bookmarks = true]{hyperref} 
\usepackage[capitalize]{cleveref} 

\usepackage{amsmath}
\usepackage{amsfonts}
\usepackage{amssymb}
\usepackage{empheq} 
\usepackage{mathtools} 

\usepackage{siunitx} 
\usepackage{braket} 

\newcommand{\avg}[1]{\left\langle #1 \right\rangle} 
\let\vaccent=\v 
\renewcommand{\v}[1]{\boldsymbol{\mathbf{#1}}} 
\newcommand{\inv}{^{-1}} 
\newcommand{\e}[1]{\, \mathrm{e}^{#1}} 
\renewcommand{\i}{\mathrm{i}} 
\newcommand{\cpi}{\uppi} 
\providecommand*{\groupU}{\mathrm{U}(1)} 
\providecommand*{\groupZ}{\mathbb{Z}_2} 
\providecommand*{\groupUZ}{\groupU \times \groupZ} 
\providecommand*{\groupU}[1]{\mathrm{U}(#1)} 

\makeatletter 
\providecommand*{\diff}
	  {\@ifnextchar^{\DIfF}{\DIfF^{}}}
  \def\DIfF^#1{%
	  \mathop{\mathrm{\mathstrut d}}%
		  \nolimits^{#1}\gobblespace}
  \def\gobblespace{%
	  \futurelet\diffarg\opspace}
  \def\opspace{%
	  \let\DiffSpace\!%
	  \ifx\diffarg(%
		  \let\DiffSpace\relax
	  \else
		  \ifx\diffarg[%
			  \let\DiffSpace\relax
		  \else
			  \ifx\diffarg\{%
				  \let\DiffSpace\relax
			  \fi\fi\fi\DiffSpace}
\makeatother

\begin{document}


\title{Phase transitions and anomalous normal state in superconductors with broken
time reversal symmetry }

\author{Troels Arnfred Bojesen}
\email{troels.bojesen@ntnu.no}
\affiliation{Department of Physics, Norwegian University of Science and Technology, NO-7491 Trondheim, Norway}
\author{Egor Babaev}
\affiliation{Department of Theoretical Physics, The Royal Institute of Technology, 10691 Stockholm, Sweden}
\affiliation{Physics Department, University of Massachusetts, Amherst, Massachusetts 01003, USA}
\author{Asle Sudb\o{}}
\affiliation{Department of Physics, Norwegian University of Science and Technology, NO-7491 Trondheim, Norway}


\date{\today}

\begin{abstract}
Using Monte Carlo simulations, we explore the phase diagram and the phase transitions in $\groupUZ$ $n$-band superconductors with spontaneously broken time-reversal symmetry (also termed $s+is$ superconductors), focusing on the three-band case. In the limit of infinite penetration length, the system under consideration can, for a certain parameter regime, have a single first order phase transition from a $\groupUZ$ broken state to a normal state due to a nontrivial interplay between $\groupU$ vortices and $\groupZ$ domain walls. This regime may also apply to multicomponent superfluids. For other parameters, when the free energy of the domain walls is low, the system undergoes a restoration of broken $\groupZ$ time reversal symmetry at temperatures lower than the temperature of the superconducting phase transition.{We show that inclusion of fluctuations can strongly suppress the temperature of the $\groupZ$-transition when frustration is weak. The main result of our paper is that}  for relatively short magnetic field penetration lengths, the system has a superconducting phase transition at a temperature lower than the temperature of the restoration of the broken $\groupZ$ symmetry. Thus, there appears a new phase which is $\groupU$-symmetric, but breaks $\groupZ$ time reversal symmetry, an anomalous dissipative (metallic) state.
\end{abstract}


\maketitle



\section{Introduction}

Superconductors and superfluids featuring condensates which can be described by several types of complex fields, so-called multicomponent superfluids and superconductors, can feature novel physics which is not seen in single-component systems. This is mainly due to the highly nontrivial interplay between the topological defects of the various components of the ordering fields. The discovery of superconductors such as the Iron Pnictides \cite{iron}, has generated much interest in multiband superconducting systems. In contrast to previously known two-band materials, iron-based superconductors may exhibit dramatically different physics due to the possibility of {frustrated} inter-band Josephson coupling originating with more than two bands crossing the Fermi-surface \cite{nagaosa,zlatko,johan3,maiti}. In systems with two bands crossing the Fermi surface, with concomitant ordering fields associated with each band, the Josephson coupling (which generically always is present and represents a singular perturbation to the case where no Josephson-coupling is present) locks the phase differences between the bands to 0 or $\cpi$. On the other hand, if one has three {or more} bands and the frustration of interband coupling is sufficiently strong, the ground state configuration may be one where interband phase-differences can differ from $0$ or $\cpi$. Consequently, such systems may feature a ground state with spontaneously broken time reversal symmetry (BTRS) \cite{nagaosa,zlatko}.This results in an overall spontaneously broken $\groupUZ$ symmetry \cite{johan3}, to be compared to the generic, single component case of just a spontaneously broken $\groupU$ symmetry. That this indeed may happen has recently been proposed for the hole-doped \ce{Ba_{$1-x$}K_{$x$}Fe2As2} pnictide superconductor \cite{maiti}. Such novel physics has also been proposed in  other classes of materials \cite{agterberg2011}, and this is a topic which is currently under intense investigation. For other scenarios of time reversal symmetry breakdown in iron-based superconductors, see Refs. \onlinecite{other1,other2}.
 
Multiband superconductors (more than two bands) with frustrated interband Josephson couplings feature several properties that are radically different from their two-band counterparts. These include I) the appearance of a massless so-called Leggett mode at the $\groupZ$ phase transition \cite{lin}, II) the appearance of new mixed phase-density collective modes in the state with broken time-reversal symmetry (BTRS) \cite{johan3,stanev,maiti,marciani} in contrast to  the \enquote{phase-only} Leggett collective mode in two-band materials \cite{leggett}, III) the appearance of (meta-)stable excitations characterized by $\mathbb{CP}^2$ topological invariants \cite{cp21,cp22,cp33}, IV) the appearance of new mechanisms for vortex viscosity \cite{silaevv}, and V) the appearance of a complex phase diagram with multiple transitions in two dimensions \cite{2d}.
 
Much of the discussion of the phase diagram of frustrated  three-band  superconductors has so far been limited to the mean-field level \cite{zlatko,maiti}. However, the iron-based materials feature relatively high $T_c$, as well as being far from the type-I regime. Furthermore, these materials feature a superconducting state which is inherently frustrated due to the sign of the interband Josephson-couplings. For these reasons, fluctuation effects in these materials may be quite significant in determining existing phases and their boundaries, even in three spatial dimensions.

In this paper, we study the phase diagram of a three-band superconductor in three spatial dimensions in the London limit, beyond the mean-field approximation. The results should apply to iron-based superconductors. A fluctuating gauge-field is also included in the description. The main findings of this work are as follows.  For sufficiently strong frustration induced by interband Josephson-coupling, the phase diagram acquires an unusual fluctuation-induced metallic state which is a precursor to the BTRS superconducting phase. This metallic state exhibits a broken $\groupZ$ time-reversal symmetry. It appears provided gauge-field fluctuations become strong enough in the model which we consider. The interpretation of this state is the same as for the corresponding state which has previously been discussed in the context of thin-film iron pnictide superconductors \cite{2d} although the phase diagram and nature of the phase transitions are different. Namely, although the state is non-superconducting, it features a persistent interband Josephson current in momentum space which breaks time reversal symmetry.

\section{Model}
The London model for an $n$-band superconductor is given by
\begin{widetext}
  \begin{equation}
F= 
\sum_{\alpha=1}^n\frac{|\psi_\alpha|^2}{2}(\nabla \theta_\alpha -e \v{A})^2 
+ \sum_{\alpha,\alpha'>\alpha } \eta_{\alpha\alpha'}|\psi_\alpha||\psi_{\alpha'}|\cos(\theta_\alpha-\theta_{\alpha'} )
+\frac{1}{2}(\nabla \times \v A)^2.
\label{eq:london}
\end{equation}
\end{widetext}
Here, $|\psi_{\alpha}| \e{\i \theta_{\alpha}}$ denote the superconducting condensate components in different bands labeled by $\alpha \in [1,..,n]$, while the second term represents interband Josephson couplings. The field $\v A$ is the magnetic vector potential that couples minimally to the charged condensate matter fields. {In this work no external magnetic field is applied.} By collecting gradient terms for phase differences, {\cref{eq:london}} can also be cast in the form 
\begin{widetext}
\begin{equation}   
F= \frac{1}{2\varrho^2} \left( \sum_\alpha |\psi_\alpha|^2\nabla \theta_\alpha -e \varrho^2{\bf A}\right)^2  +  \frac{1}{2}(\nabla \times {\bf A})^2+ 
\sum_{\alpha,\alpha'>\alpha}\frac{|\psi_\alpha|^2|\psi_{\alpha'}|^2}{2\varrho^2}[\nabla(\theta_\alpha-\theta_{\alpha'} )]^2
 +\eta_{\alpha\alpha'}|\psi_\alpha||\psi_{\alpha'}|\cos(\theta_\alpha-\theta_{\alpha'} ),
\label{eq:ps}		
\end{equation}
\end{widetext}
where $\varrho^2=\sum_\alpha |\psi_\alpha|^2$. Thus, the vector potential is coupled to the $\groupU$ sector of the model, but not to phase differences.

\begin{figure}[ht]
  \subfloat[Phases of the fields.\label{fig:phase_definition}]{
    \def\svgwidth{0.4\columnwidth}
  \begingroup%
  \makeatletter%
  \providecommand\color[2][]{%
    \errmessage{(Inkscape) Color is used for the text in Inkscape, but the package 'color.sty' is not loaded}%
    \renewcommand\color[2][]{}%
  }%
  \providecommand\transparent[1]{%
    \errmessage{(Inkscape) Transparency is used (non-zero) for the text in Inkscape, but the package 'transparent.sty' is not loaded}%
    \renewcommand\transparent[1]{}%
  }%
  \providecommand\rotatebox[2]{#2}%
  \ifx\svgwidth\undefined%
    \setlength{\unitlength}{135.16518555bp}%
    \ifx\svgscale\undefined%
      \relax%
    \else%
      \setlength{\unitlength}{\unitlength * \real{\svgscale}}%
    \fi%
  \else%
    \setlength{\unitlength}{\svgwidth}%
  \fi%
  \global\let\svgwidth\undefined%
  \global\let\svgscale\undefined%
  \makeatother%
  \begin{picture}(1,0.59557126)%
    \put(0,0){\includegraphics[width=\unitlength]{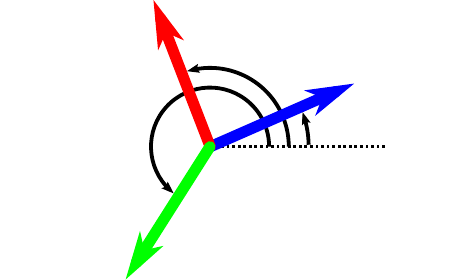}}%
    \put(0.68813623,0.30221329){\color[rgb]{0,0,0}\makebox(0,0)[lb]{\smash{$\theta_{1}$}}}%
    \put(0.5098767,0.4595013){\color[rgb]{0,0,0}\makebox(0,0)[lb]{\smash{$\theta_{2}$}}}%
    \put(0.31188689,0.31269923){\color[rgb]{0,0,0}\makebox(0,0)[rb]{\smash{$\theta_{3}$}}}%
  \end{picture}%
  \endgroup%
  }\\
  \subfloat[$+1$]{\includegraphics[width=0.4\columnwidth]{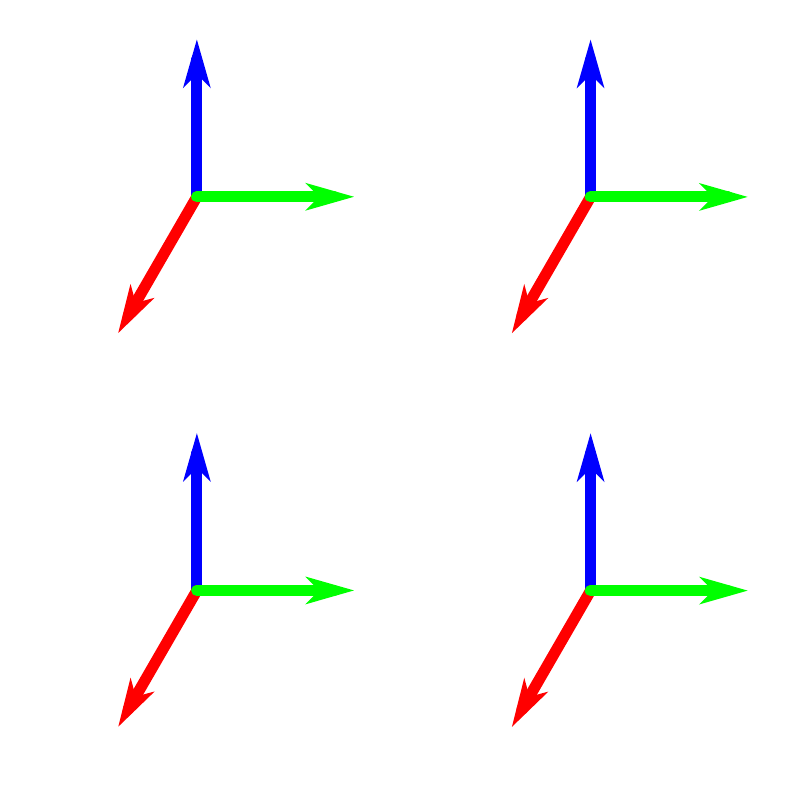}}
  \qquad
  \subfloat[$-1$]{\includegraphics[width=0.4\columnwidth]{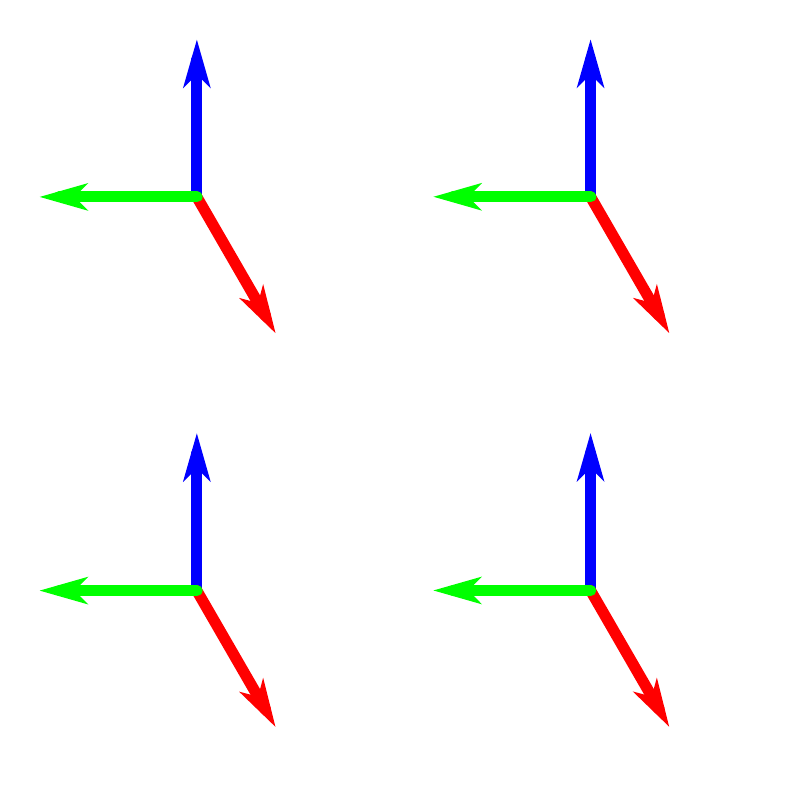}}
  \caption{(Colors online) The arrows in panel a) $(\textcolor{blue}{\longrightarrow},\textcolor{red}{\longrightarrow},\textcolor{green}{\longrightarrow})$ correspond to $(\theta_1,\theta_2,\theta_3)$. Panels (b) and (c) show examples of phase configurations for the two $\groupZ$ symmetry classes of the ground states, shown on a $2\times 2$ lattice of selected points of a planar slice of the system. Here $g_{12} > g_{23} > g_{13} > 0$. The spatial contribution to the energy is minimized by making the spatial gradient zero (hence breaking the global $\groupU$ symmetry). Then there are two classes of phase configurations, one with chirality +1 and one with chirality -1, minimizing the energy associated with the interband interaction. The chirality is defined as $+1$ if the phases (modulo $2\cpi$) are cyclically ordered $\theta_1<\theta_2<\theta_3$, and $-1$ if not.}  
\label{fig:gr_state_Z2}
\end{figure}

When the Josephson couplings $\eta_{\alpha\alpha'}$ are positive, each Josephson term by itself prefers to lock phase differences to $\cpi$, i.e. $\theta_\alpha-\theta_{\alpha'} =\cpi $. Since this is not possible for three phases or more, the system is generically frustrated. {For certain parameter values,} the system breaks time reversal symmetry when Josephson couplings are minimized by two inequivalent phase lockings, shown in Fig. \ref{fig:gr_state_Z2} {for the three band case}. The phase lockings are related by complex conjugation of the fields $\psi_\alpha$. Thus, by choosing one of these phase locking patterns the system breaks time reversal symmetry \cite{nagaosa,zlatko,johan3} Note also that  there are special cases where the degeneracy is larger \cite{Weston_2013}, but they have measure zero in phase space of the model in question and are ignored here.
{This model} 
{describes a $s+\i s$ superconductor in the London limit. For the parameters where the model breaks $\groupUZ$ symmetry it allows topological excitations in the form of domain walls as well as composite vortices. In the composite vortices all the phases wind by $2\cpi$ and thus they do not carry a topological charge in the $\groupZ$ sector. Thus proliferation of such vortices cannot disorder phase difference and therefore the system can in principle have a state with broken $\groupZ$ symmetry but with restored $\groupU$ symmetry. Since in this model there is also nontrivial interaction between the vortices and the domain walls it requires careful numerical examination under what conditions such a phase may occur (for detailed discussion of vortex and domain wall solutions and their interaction see Ref. \onlinecite{cp22}).}

\subsection{Lattice model}

The lattice version of \cref{eq:london} reads
\begin{multline}
H = - \sum_{\avg{i,j},\alpha} ~ a_\alpha ~\cos\left(\theta_{\alpha,i} - \theta_{\alpha,j} - A_{ij}\right) \\
+ \sum_{i,\alpha' > \alpha} g_{\alpha \alpha'} \cos\left(\theta_{\alpha,i} - \theta_{\alpha',i} \right) \\
+ q \sum_{i,\lambda} \Bigl(\sum_{\mu,\nu} \epsilon_{\lambda \mu \nu}\Delta_{\mu}A_{i,i+\nu}\Bigr)^2.
\label{eq:H_gauge}
\end{multline}
Here, $i,j \in \set{1,2,\ldots,N=L^3}$ denote sites on a lattice of size $L \times L \times L$, and $\avg{i,j}$ indicates pairs of nearest neighbor lattice sites (assuming periodic boundary conditions). We may, without loss of generality, choose
\begin{equation}
 a_{1} = 1, \qquad a_{\alpha} \in (0,1], \quad \alpha > 1,
 \label{eq:a_restriction}
\end{equation}
where $g_{\alpha \alpha'}$ are interband Josephson couplings. We have have rescaled the gauge field $\v A \gets e\v A$ and introduced 
\begin{equation}
 q \equiv 1/(2e^2).
 \label{eq:q_def}
\end{equation}
{In these units $q$ parametrizes the London penetration depth of the superconductor.}

In the limit $e \to 0 \Leftrightarrow q \to \infty$, where fluctuations in the gauge field may be neglected, the model is reduced to
\begin{multline}
H = - \sum_{\avg{i,j},\alpha} a_\alpha \cos\left(\theta_{\alpha,i} - \theta_{\alpha,j} \right) \\
+ \sum_{i,\alpha' > \alpha} g_{\alpha \alpha'} \cos\left(\theta_{\alpha,i} - \theta_{\alpha',i} \right).
\label{eq:H_basic}
\end{multline}

By letting $g_{\alpha \alpha'} \to \infty$ {in the lattice London model} such that the ratio $g_{\alpha \alpha'}/g_{\beta\beta'}$ is finite, we may derive a \enquote{reduced} version of the model given by \cref{eq:H_gauge,eq:H_basic}, for which the intercomponent phase fluctuations effectively are neglected. Namely, the \enquote{phase star} of a lattice site locks into one of the two possible $\groupZ$ configurations minimizing the contribution from the Josephson term in the Hamiltonian. That is, in this approximation the phase differences can have only two values. The $\groupZ$ domain wall then represents a change of the phase difference at one lattice spacing.

The reduced {lattice London model} is given by a rather unusual coupled Ising-XY type of model, 
\begin{multline}
 H = -\sum_{\langle i,j\rangle}\big[ (1 + K_1\sigma_i \sigma_j)\cos(\theta_i - \theta_j - A_{ij}) \\
 + K_2(\sigma_i - \sigma_j)\sin(\theta_i - \theta_j - A_{ij}) \big] \\
+ q \sum_{i,\lambda} \Bigl(\sum_{\mu,\nu} \epsilon_{\lambda \mu \nu}\Delta_{\mu}A_{i,i+\nu}\Bigr)^2,
 \label{eq:H_reduced_gauge}
\end{multline}
and
\begin{multline}
 H = -\sum_{\langle i,j\rangle}\big[ (1 + K_1\sigma_i \sigma_j)\cos(\theta_i - \theta_j) \\
 + K_2(\sigma_i - \sigma_j)\sin(\theta_i - \theta_j) \big],
 \label{eq:H_reduced}
\end{multline}
for the cases with and without fluctuating gauge-field. Here $\sigma \in \set{-1,+1}$ denotes the $\groupZ$ chirality of the \enquote{phase star}, $\theta_i \equiv \theta_{1,i}$, and
\begin{align}
 K_1 \equiv& \frac{\sum_{\alpha>1} a_\alpha\bigl[1-\cos(2\phi_{\alpha})\bigr]}{2 + \sum_{\alpha>1} a_\alpha \bigl[1+\cos(2\phi_{\alpha})\bigr]} \label{eq:K1} \\
 K_2 \equiv& \frac{\sum_{\alpha>1} a_\alpha \sin(2\phi_{\alpha})}{2 + \sum_{\alpha>1} a_\alpha \bigl[1+\cos(2\phi_{\alpha})\bigr]}.
 \label{eq:K2}
\end{align}
The angles $\phi_\alpha \equiv \theta_{\alpha,i}-\theta_{1,i}, \alpha > 1$ are determined by the ratios $g_{\alpha \alpha'}/g_{\beta\beta'}$ of the Josephson-couplings. For site-independent Josephson-couplings, the $\phi_\alpha$'s are also site-independent. See \cref{app:K1K2_model} for details.

The motivation for introducing a {reduced lattice London model} is that it is much simpler than the model in \cref{eq:H_gauge}, while it appears to exhibit much of the same physics, at least at the level of the phase diagram. 

A remark about taking the limit $g_{\alpha \alpha'} \to \infty$ is nevertheless in order here, since it eliminates gradients in intercomponent phase differences.  
Although such an approximation is trivial in the non-frustrated case, it is more subtle in the $\groupUZ$ case. 
{First of all, note that we take this limit only in the lattice model.
In the lattice version of this model, there is a finite energy cost associated with $\groupZ$ domain walls because the lattice provides minimal
length scale of the theory. The lattice system therefore features entropy-driven proliferation of $\groupZ$ domain walls at a finite certain temperature.} As we will see below, a $\groupZ$ phase transition can take place at the same temperature or close to a $\groupU$ phase transition. This implies the coexistence of thermally induced composite vortices and domain walls. In general, their interaction produces fractional vortices \cite{cp21,cp22,cp33} even if the Josephson coupling is strong. 

An unusual feature of \cref{eq:H_reduced_gauge} is the appearance of the $K_2$-term, which favors states with large phase-differences $\theta_i - \theta_j$ on the links of the lattice. 

The domain of $(K_1,K_2)$ is given by the filled ellipse
\begin{equation}
 \left[\frac{2}{n-1}K_1 - 1\right]^2 + \left[\frac{2\sqrt{n}}{n-1}K_2\right]^2 \leq 1,
 \label{eq:K1K2_domain}
\end{equation}
as shown in \cref{app:K1K2_domain}. In \cref{app:staggered_criterion} we show that within the $n$-band model {(\ref{eq:london})}, the $K_1$ term will always dominate over the $K_2$ term in the ground state, preventing the formation of a $K_2$ dominated staggered flux phase ground state. Thus, to the extent that we will study this model in the present paper, we focus on the case $K_2=0$. It should however be noted that the model given by \cref{eq:H_reduced_gauge} is interesting in its own right as a model in statistical physics featuring phase transitions between uniform and textured ground states. In \cref{app:symmetries} we consider the symmetries of the model in $(K_1,K_2)$-space.

\section{Simulation Results}

We have performed extensive Monte-Carlo simulations of the models defined by \cref{eq:H_gauge,eq:H_basic} for the case $n=3$ and $a_{\alpha} = 1 \quad \forall \alpha$, that is both with and without a fluctuating gauge-field in the problem. For technical details pertaining to our approach, see \cref{app:microcanonical,numerical_techniques}. We have also considered the reduced case with infinite Josephson-couplings with fixed ratios, defined by \cref{eq:H_reduced_gauge,eq:H_reduced} with $n=3$, for the case $K_2=0$. The results obtained for $K_2=0$ are expected to hold also for $K_2 \ll 1$. 

\begin{figure}[ht]
  \includegraphics{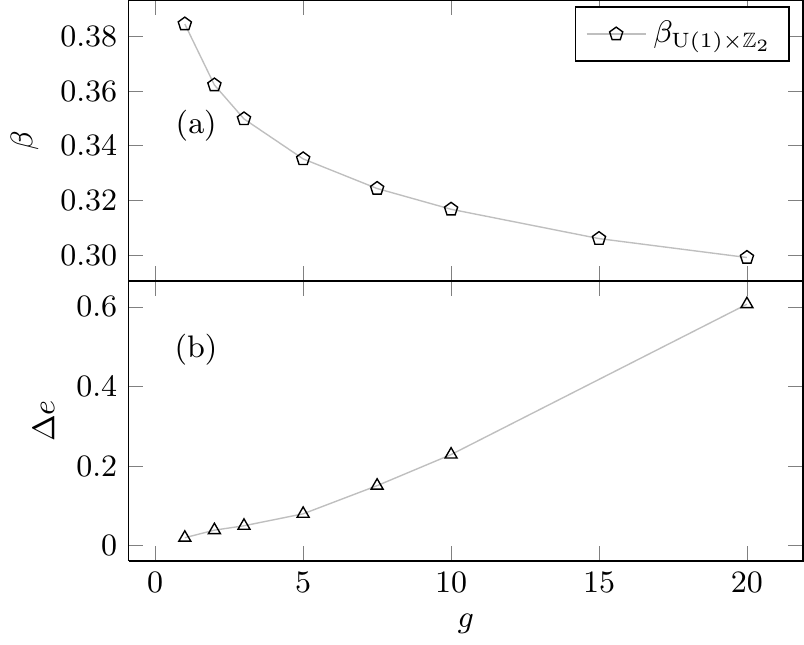}
\caption{(a) The phase diagram of the full model without gauge field, \cref{eq:H_basic}, when $g_{12} = g_{23} = g_{31} = g$. Both symmetries, $\groupU$ and $\groupZ$, experience a simultaneous first order phase transition when crossing the $\beta_{\groupUZ}$ line. In the limit $g\to\infty$ (which corresponds to \cref{eq:H_reduced} with $(K_1,K_2)=(1,0)$) $\beta_{\groupUZ} \to 0.2729$. {For $g = 0$, the model reduces to the triply degenerate 3D XY} model with a second order phase transition at $\beta_\text{XY} = \num{0.4541674(1)}$~\cite{2012arXiv1211.0780L}. (b) The latent heat $\Delta e$ of the transition as a function of $g$, clearly showing that the first order character of the transition becomes stronger as $g$ increases. In the limit $g\to\infty$ (not shown), $\Delta e \to 1.071$. Results for $g=1,2,3$ are based on simulations with $L=120$, results for $g=5,7.5,10,15,20$ on simulations with $L=80$, while the results for $g\to\infty$ are based on a simulation with $L=50$. 
\label{fig:full_nogauge_g}}
\end{figure}

Since the Josephson couplings $\set{g}$ represent singular perturbations in the Hamiltonians given by \cref{eq:H_gauge,eq:H_basic}, which explicitly break $\groupU\times \groupU \times \groupU$ symmetry down to $\groupUZ$, one may {naively} expect the qualitative behavior of the models to be similar at least for strong coupling $\set{g}$. \Cref{fig:full_nogauge_g} shows that this, within the parameter regime we have been able to access, indeed is the case. For these parameters the model exhibits a first order transition from a $\groupUZ$ broken state to a symmetric state. The transition does not change its character in the range $g \in [1,\infty)$. {Simulating the phase diagram in the limit $g \ll 1$ is computationally extremely demanding.  In this limit, the width of the domain walls grows due to growing Josephson length. Thus, it requires much larger lattice sizes to observe a  splitting of  the $\groupZ$ and  $\groupU$ phase transitions, a splitting which is suggested by  the decreasing energy of domain walls in small-$g$ limit. A study of this limit is beyond the scope of this work. Nonetheless, in this regime we have performed extra Monte Carlo simulations using a different algorithm, as explained in \cref{app:new_numerical}. For couplings as small as $g=0.001$, we were unable to detect any splitting, even for a system size of $L=128$. However, the above mentioned growth of domain-wall widths will make this splitting indetectable in finite systems in the limit $g\approx 0$. This calls  for further investigation of this limit.}

\begin{figure}
\includegraphics{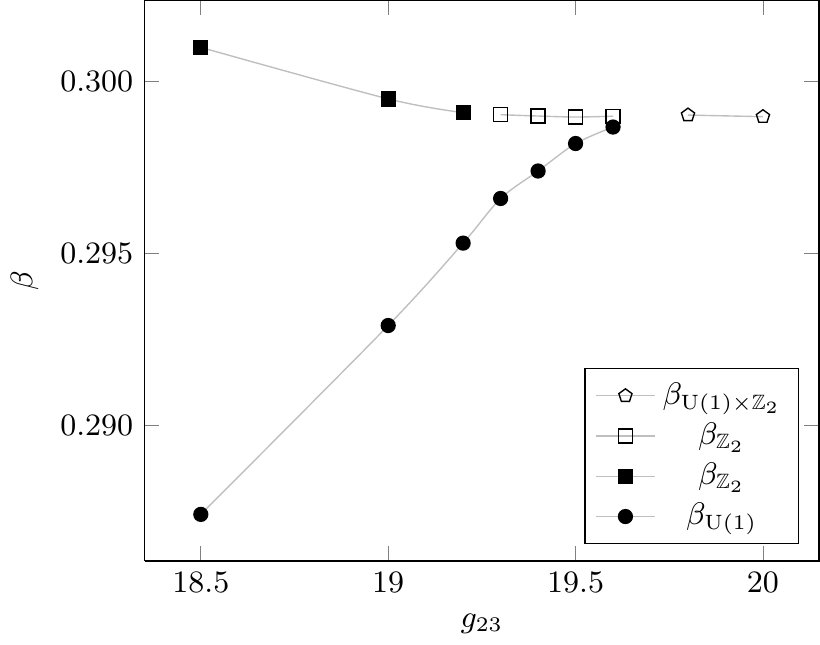}
\caption{The phase diagram of the full model without gauge field, \cref{eq:H_basic}, as a function of $g_{23}$. We have fixed $g_{12} = g_{13} = 20$. Open symbols indicate first order phase transitions, while filled symbols indicate second order transitions. The symmetries involved are shown as subscripts of the $\beta$'s in the legend. The data are based on simulations with $L=40$. Error bars are smaller than symbol sizes, except for $\beta_{\groupU}$ where they are of comparable size. To be compared with \cref{fig:reduced_nogauge_K}. 
\label{fig:full_nogauge_g23}} 
\end{figure}

\begin{figure}
\includegraphics{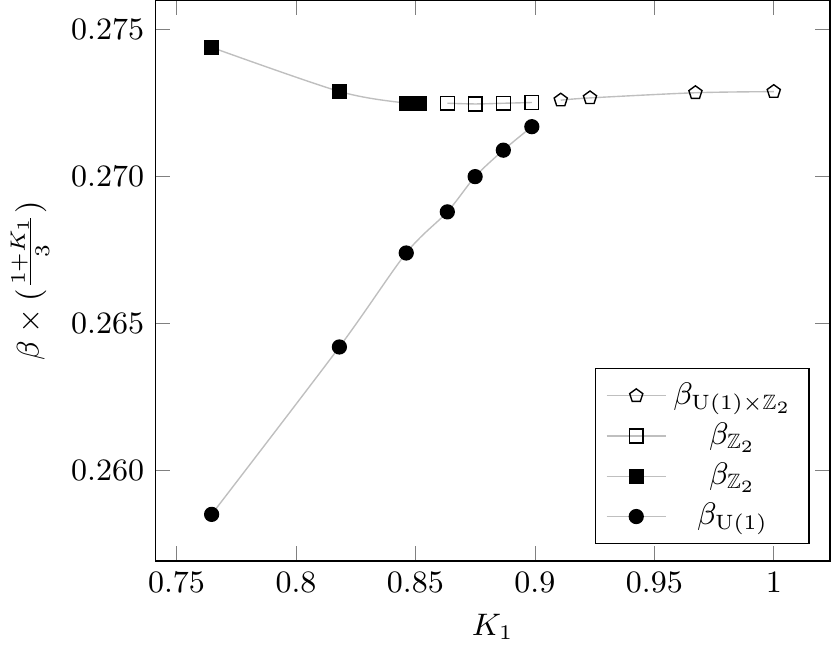}
\caption{The phase diagram of the reduced model without gauge field, \cref{eq:H_reduced}, as a function of $K_1$. $K_2 = 0$. Open symbols indicate first order phase transitions, while filled symbols indicate second order transitions. The symmetries involved are shown as subscripts of the $\beta$'s in the legend. The data are based on simulations with $L=50$. Errors bars are smaller than symbol sizes, except for $\beta_{\groupU}$ where they are of comparable size. {The ordinate is scaled by $(1+K_1)/3$ to make it coincide with the full model in the limit $\set{g\to\infty}$.} To be compared with \cref{fig:full_nogauge_g23}. 
\label{fig:reduced_nogauge_K}}
\end{figure}

Consider next the effect of varying the ratios of the Josephson couplings. \Cref{fig:full_nogauge_g23,fig:reduced_nogauge_K} show the phase diagrams of the full model, \cref{eq:H_basic}, and the reduced model, \cref{eq:H_reduced}, i.e. there is no fluctuating gauge field in the problem. Here, the anisotropy of the ground state \enquote{phase star} is varied, starting with maximal symmetry when $g_{23} = 20$ or $K_1 = 1$. Near the maximally symmetric ground states, there is a single first order phase transition from a fully ordered state with broken $\groupUZ$ symmetry to a fully disordered state which is $\groupUZ$-symmetric. That is, the $\groupU$- and $\groupZ$ sectors do not order independently. Increasing the anisotropy by letting $g_{23}$ or $K_1$ decrease, the $\groupU$ sector order becomes less affected by the disorder of the $\groupZ$ sector.  This eventually leads to a second order $\groupU$ transition line bifurcating from the $\beta_{\groupUZ}$ curve, leaving a first order $\beta_{\groupZ}$ transition line. As the anisotropy increases, the coupling between the symmetry sectors becomes small enough for the fluctuations in the broken $\groupU$ to be insignificant for the $\groupZ$ transition. The $\groupZ$ transition then goes through a tricritical point and reaches its second order $\groupZ$ universality class nature. In the maximally anisotropic limit ($g_{23} \leq 10$\footnote{The ground state contribution of the Josephson term, i.e. the minimum of $\sum_{\alpha>\alpha'} g_{\alpha\alpha'}\cos(\theta_\alpha - \theta_{\alpha'})$, is reached when $\theta_2 = \theta_3$ when $g_{23} \leq g/2$, when $g_{12} = g_{13} = g$.} or $K_1\to 0$) the symmetry  of the model is explicitly broken further down from $\groupUZ$ to just $\groupU$; it is no longer possible to distinguish between $\sigma = +1$ and $\sigma = -1$ fields, nor is there an energy barrier between them. Then $\beta_{\groupZ}\to \infty$, and the model is effectively reduced to the ordinary 3D XY model.

It is possible to switch the positions of $\groupZ$ and $\groupU$ transition in the phase diagrams, meaning that the $\groupZ$-symmetry is broken before the $\groupU$-symmetry upon cooling the system from the disordered side. This takes place for  sufficiently strong coupling to the gauge field in the models, \cref{eq:H_gauge,eq:H_reduced_gauge}. The result of varying the coupling is shown in \cref{fig:reduced_gauge_q,fig:full_gauge_q}. For large values of $q$, i.e. small values of the charge $e$ (see \cref{eq:q_def}), the overall structure of the phase diagram is similar to the case of no gauge field at all. (Note that the ground state phase stars are maximally symmetric here.) As $q$ is decreased, the gauge field fluctuations increase, allowing the $\groupU$ order parameter to disorder at lower temperatures without significantly affecting the $\groupZ$ order. This is because the characteristic energy scale associated with creating large vortex loops is much lower than the characteristic energy scale associated with proliferating $\groupZ$ domain walls. The result is a splitting of the first order $\beta_{\groupUZ}$ line into two second order $\beta_{\groupZ}$- and $\beta_{\groupU}$ lines. {Note that, unlike the cases shown in Figs. \ref{fig:full_nogauge_g23} and 
\ref{fig:reduced_nogauge_K}, the splittings shown in Figs. \ref{fig:full_gauge_q} and \ref{fig:reduced_gauge_q} show no signs of a first-order transition after the splitting point
within the resolution of our simulations. {It does not exclude the possibility that it requires larger lattices to detect the first order phase transition for some part of the line
when $\beta_{\groupU} > \beta_{\groupZ}$.}

\begin{figure}
\includegraphics{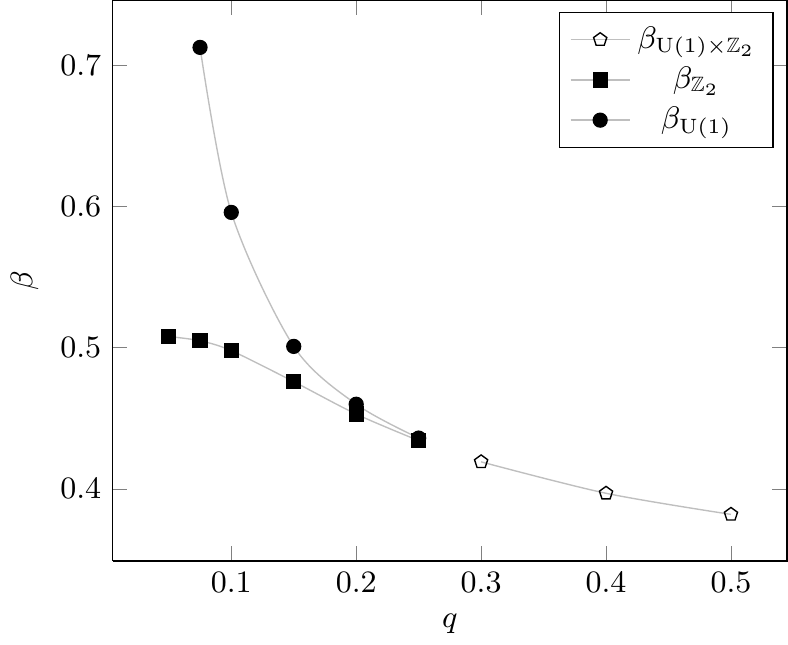}
\caption{The phase diagram of the full model, \cref{eq:H_gauge}, as a function of $q = 1/2e^2$. We have $g_{12} = g_{13} = g_{23} = 20$. Open symbols indicate first order phase transitions, filled symbols indicate second order transitions. The symmetries involved are shown as subscripts of the $\beta$'s in the legend. The data are based on simulations with $L=40$. Error bars are smaller than symbol sizes, except for $\beta_{\groupU}$ where they are of comparable size. To be compared with \cref{fig:reduced_gauge_q}.
\label{fig:full_gauge_q}}
\end{figure}

\begin{figure}
\includegraphics{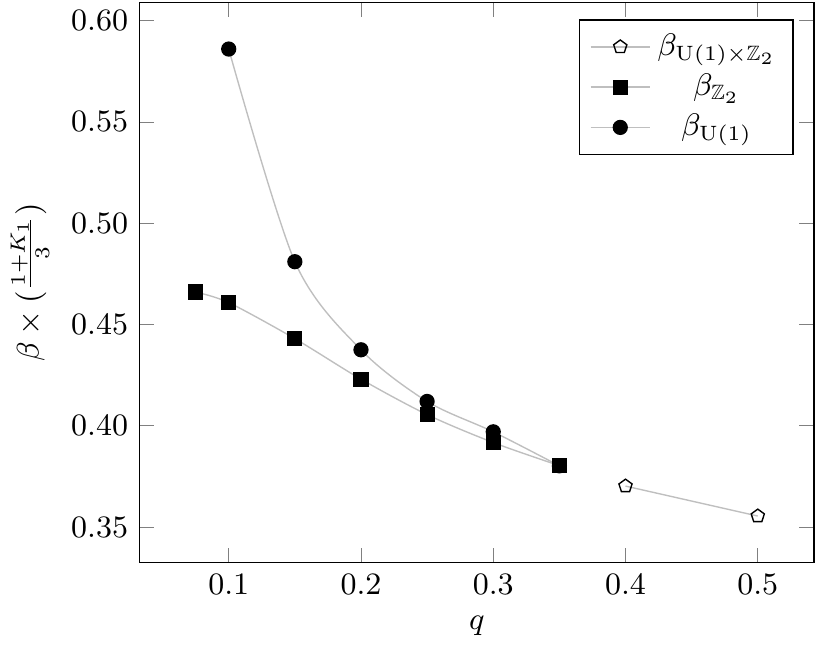}
\caption{The phase diagram of the reduced model, \cref{eq:H_reduced_gauge}, as a function of $q = 1/2e^2 $. We have $K_1=1$, while $K_2=0$. Open symbols indicate first order transitions; filled symbols indicate second order transitions. The symmetries involved are shown as subscripts of the $\beta$'s in the legend. The data are based on  simulations with $L=40$. Error bars are smaller than symbol sizes, except for $\beta_{\groupU}$ where they are of comparable size. {The ordinate is scaled by $(1+K_1)/3$ to make it coincide with the full model in the $\set{g\to\infty}$ limit.} To be compared with \cref{fig:full_gauge_q}.
\label{fig:reduced_gauge_q}}
\end{figure}

We have identified two situations where the $\beta_{\groupU}$ and $\beta_{\groupZ}$ transition lines are separate. Namely, one with $\beta_{\groupZ} > \beta_{\groupU}$, as in \cref{fig:full_nogauge_g23,fig:reduced_nogauge_K} for larger anisotropies, but no gauge field (effectively $q\to\infty$), and one with $\beta_{\groupU} > \beta_{\groupZ}$, as in \cref{fig:full_gauge_q,fig:reduced_gauge_q} for small values of $q$. Consider now the case where these tendencies compete. A plot illustrating this is shown in \cref{fig:full_gauge_g23}. Within the resolution of our simulations, the lines seem to cross in a single point, although we cannot exclude the possibility of a short segment where the lines merge forming a single first order phase transition line.

\begin{figure}
\includegraphics{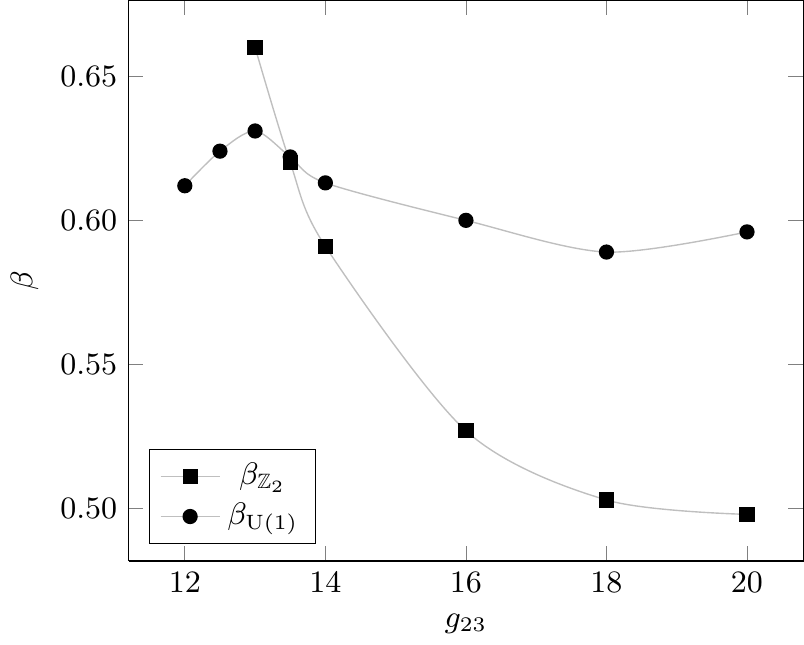}
\caption{The phase diagram of the full model, \cref{eq:H_gauge}, as a function of $g_{23}$. $g_{12} = g_{13} = 20$ and $q = 0.1$. The symmetries involved in the transitions are shown as subscripts of the $\beta$'s in the legend. Both the $\groupU$ and the $\groupZ$ transitions are of second order {within the resolution of our simulations}. The data are based on  simulations with $L=40$.  \label{fig:full_gauge_g23}} 
\end{figure}

\section{Discussion}
We have discussed the phase diagram of multiband (more than two bands) superconductors exhibiting spontaneously broken time reversal symmetry. Determining the phase diagram of such systems beyond a mean field approximation is highly non-trivial, due to a delicate interplay between the thermally excited topological objects of the systems, which are $\groupZ$ (Ising) domain walls, and $\groupU$ vortex loops. 

The central result of this work is that we have shown that for relatively short coherence lengths, a three dimensional $\groupUZ$ superconductor/superfluid can feature an anomalous non-superconducting/non-superfluid state, in which time-reversal symmetry has been spontaneously broken. This is a novel phase which is not found at the mean-field level; it can only be found by taking fully into account the critical fluctuations of the system. In this state, the system retains order in the phase differences of the various components of the ordering field. Thus, it should feature persistent intercomponent (interband) currents in $\v k$-space. Experimental verification of this state would require a probe of the phase difference excitations (e.g. local tunneling probes or detection of the phase-difference mixed collective modes \cite{johan3,stanev,maiti}) concomitant with dissipative transport properties. In addition, a strong indication of such an anomalous state would be the detection of a 3D Ising anomaly in the specific heat above the superconducting transition temperature, i.e. inside the non-superconducing/non-superfluid state.

We have demonstrated that under certain conditions the interactions between $\groupZ$ Ising domain walls and $\groupU$ vortex loops { makes a direct $\groupUZ$ superconductor/superfluid-normal metal/normal fluid transition first order, in contrast to the corresponding transition in $\groupU$ type-II superconductors or superfluids.}

{The overall structure of the phase diagram for the 3D system is quite different from the phase diagram of the same system in 2D \cite{2d}. In particular, in 2D there does not exist a large parameter regime where the $\groupZ$ and $\groupU$ phase transitions merge into a single first order phase transition.  The existence of such a regime in 3D is due to a preemptive effect whereby vortex loops assist the proliferation of $\groupZ$ domain walls and vice versa via a formation of composite defects which
carry both $\groupZ$ and $\groupU$ topological charges (for a description of composite defects see \cite{cp22}) .
 In 2D, the corresponding phenomenon would be due to a dilute system of vortices and antivortices  assisting the proliferation of $\groupZ$ domain lines. } The absence and  presence of a preemptive first order transition in 2D and 3D, respectively, shows that thermal creation of $\groupU$ topological excitations near the superconducting phase transition assists the disordering of the $\groupZ$ sector to a lesser degree in 2D compared to the 3D case. 

Finally, we mention that we expect the results we have obtained for three-band systems to hold also for other systems which break time reversal symmetry. This is seen to be the case for two-component systems with fourth or higher-order intercomponent coupling of the type $\psi_1^2\psi_2^*{}^2 +c.c.$, when we note that the reduced version of such a model is the $K_1 K_2$ model with $K_2 = 0$. Moreover, we expect our results to hold also for frustrated superconductors with four- and larger number of components. {Namely, it can be shown that when the number of components exceeds $n=3$, a $\groupUZ$ symmetry nonetheless emerges}\cite{Weston_2013}.

\begin{acknowledgments}
T.A.B. thanks NTNU for financial support, and the Norwegian consortium for high-performance computing (NOTUR) for computer time and technical support.
A.S. was supported by the Research Council of Norway, through Grants 205591/V20 and 216700/F20. E.B. was supported by Knut and Alice Wallenberg Foundation 
through the Royal Swedish Academy of Sciences Fellowship, Swedish Research Council and by the National Science Foundation CAREER Award No. DMR-0955902.
\end{acknowledgments}

\appendix

\section{Derivation of the reduced $\groupUZ$ model, \cref{eq:H_reduced_gauge,eq:H_reduced} \label{app:K1K2_model}}
Starting from \cref{eq:H_gauge,eq:H_basic}, we derive the reduced model given by \cref{eq:H_reduced_gauge,eq:H_reduced} by letting $g_{\alpha \alpha'} \to \infty$ such that the ratio $g_{\alpha \alpha'}/g_{\beta\beta'}$ is kept finite. In this way the \enquote{phase star} locks to one of the two configurations minimizing the Josephson term in the Hamiltonian. The intercomponent fluctuations are thus eliminated, and the \enquote{phase star} of a lattice site may be completely determined by an overall $\groupU$ phase, $\theta$, a $\groupZ$ chirality order parameter, $\sigma$, and the (constant) positive angle between $\theta_{1}$ and $\theta_{\alpha}$, $\phi_\alpha$, as shown in \cref{fig:K1K2_derivation}. Since $g_{\alpha\alpha'} > 0$, the set of possible $\set{\phi_\alpha}$'s minimizing the Josephson term of \cref{eq:H_gauge} must be such that the phase vectors (when restricting to positive vector directions) span more than a half-plane. The $n-1$ other degrees of freedom in the choice of $\set{\phi_\alpha}$ are determined by the set of Josephson couplings $\set{g_{\alpha\alpha'}}$.

{The phase differences $\phi_\alpha$ do not couple to the gauge field. We thus first derive the reduced model without a gauge field \cref{eq:H_reduced}. The reduced model including a fluctuating gauge field, \cref{eq:H_reduced_gauge}, is then obtained by replacing $\theta_i - \theta_j \to \theta_i - \theta_j - A_{ij}$ and adding a Maxwell term.}

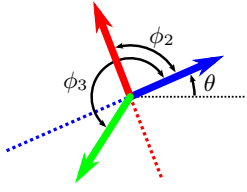
\begin{figure}
  \def\svgwidth{0.4\columnwidth}
  \begingroup%
    \makeatletter%
    \providecommand\color[2][]{%
      \errmessage{(Inkscape) Color is used for the text in Inkscape, but the package 'color.sty' is not loaded}%
      \renewcommand\color[2][]{}%
    }%
    \providecommand\transparent[1]{%
      \errmessage{(Inkscape) Transparency is used (non-zero) for the text in Inkscape, but the package 'transparent.sty' is not loaded}%
      \renewcommand\transparent[1]{}%
    }%
    \providecommand\rotatebox[2]{#2}%
    \ifx\svgwidth\undefined%
      \setlength{\unitlength}{114.7145525bp}%
      \ifx\svgscale\undefined%
	\relax%
      \else%
	\setlength{\unitlength}{\unitlength * \real{\svgscale}}%
      \fi%
    \else%
      \setlength{\unitlength}{\svgwidth}%
    \fi%
    \global\let\svgwidth\undefined%
    \global\let\svgscale\undefined%
    \makeatother%
    \begin{picture}(1,0.70174618)%
      \put(0,0){\includegraphics[width=\unitlength]{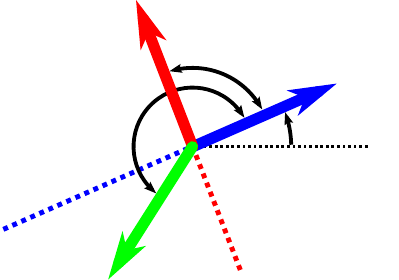}}%
      \put(0.76700319,0.35609009){\color[rgb]{0,0,0}\makebox(0,0)[lb]{\smash{$\theta$}}}%
      \put(0.5569646,0.54141848){\color[rgb]{0,0,0}\makebox(0,0)[lb]{\smash{$\phi_{2}$}}}%
      \put(0.32367835,0.3684454){\color[rgb]{0,0,0}\makebox(0,0)[rb]{\smash{$\phi_{3}$}}}%
    \end{picture}%
  \endgroup%
  \caption{(Colors online) One of the two $\groupZ$ phase configurations in the $g_{\alpha\alpha'}\to \infty$ limit when $n=3$. The dashed lines indicate the restrictions on $\phi_3$; the phase vector cannot be outside this sector.}
\label{fig:K1K2_derivation}
\end{figure}

We let $\theta = \theta_1$ represent the overall angle of the \enquote{phase star}, and label the two $\groupZ$ configurations $\sigma = 1$ and $\sigma = -1$. One configuration is obtained from the other by mirroring the phases about the axis spanned by the $\theta$ phase vector.

Links between two neighboring lattice sites $i$ and $j$ can now be divided into two categories, namely \enquote{ferromagnetic} (FM) when $\sigma_i = \sigma_j$, and \enquote{antiferromagnetic} (AFM) when $\sigma_i = -\sigma_j$. For FM links, the contribution to the Hamiltonian is simply given by
\begin{equation}
 H_{ij} = -\sum_{\alpha}a_{\alpha}\cos(\theta_i - \theta_j).
\end{equation}

For AFM links, and when $\sigma_i = 1 = -\sigma_j$, the contribution to the Hamiltonian is given by
\begin{equation}
\begin{split}
  H_{ij} &= - \sum_{\alpha}a_{\alpha}\cos((\theta_i - \phi_{\alpha}) - (\theta_j + \phi_{\alpha})) \\
  &= - \sum_{\alpha}a_{\alpha}\cos(\theta_i - \theta_j - 2\phi_{\alpha}) \\
  &= - \sum_{\alpha}a_{\alpha}\bigl[ \cos(2\phi_{\alpha})\cos(\theta_i - \theta_j ) \\
  &\hphantom{{}= - \sum_{\alpha}a_{\alpha}} {}+ \sin(2\phi_{\alpha})\sin(\theta_i - \theta_j)\bigr].
\end{split}
\label{eq:hij_anti1}
\end{equation}
For AFM links, and when $\sigma_i = -1 = -\sigma_j$, the contribution to the Hamiltonian is given by
\begin{equation}
\begin{split}
  H_{ij} &= - \sum_{\alpha}a_{\alpha}\bigl[ \cos(-2\phi_{\alpha})\cos(\theta_i - \theta_j ) \\
  &\hphantom{{}= - \sum_{\alpha}a_{\alpha}} {}+ \sin(-2\phi_{\alpha})\sin(\theta_i - \theta_j)\bigr],
\end{split}
\label{eq:hij_anti2}
\end{equation}
By using $\sigma_i$, \cref{eq:hij_anti1,eq:hij_anti2} may be combined to a single expression for the AFM links
\begin{equation}
\begin{split}
  H_{ij} &= - \sum_{\alpha}a_{\alpha}\bigl[ \cos(2\phi_{\alpha})\cos(\theta_i - \theta_j ) \\
  &\hphantom{{}= - \sum_{\alpha}a_{\alpha}} {}+ \sigma_i\sin(2\phi_{\alpha})\sin(\theta_i - \theta_j)\bigr].
\end{split}
\end{equation}

Collecting these results, and using that a Kronecker delta may be written $\delta_{\sigma_i,\sigma_j} = (1 + \sigma_i \sigma_j)/2$ and $\sigma_i^2 = 1$, we obtain
\begin{multline}
 H_{ij} = -(J_1 + J_2\sigma_i \sigma_j)\cos(\theta_i - \theta_j) \\
 - J_3(\sigma_i - \sigma_j) \sin(\theta_i - \theta_j)
 \label{eq:Hij_intermediate}
\end{multline}
where
\begin{align}
 J_1 \equiv& 1 + \sum_{\alpha>1} \frac{a_\alpha}{2}\bigl[1+\cos(2\phi_{\alpha})\bigr] \\
 J_2 \equiv& \sum_{\alpha>1} \frac{a_\alpha}{2}\bigl[1-\cos(2\phi_{\alpha})\bigr] \\
 J_3 \equiv& \sum_{\alpha>1} \frac{a_\alpha}{2}\sin(2\phi_{\alpha})
\end{align}
In \cref{eq:Hij_intermediate}, we have an overall scaling factor, $J_1$ say, which we may divide out without loss of generality. We then obtain
\begin{align}
 H &= \sum_{\langle i,j\rangle} H_{ij} \\
  &= -\sum_{\langle i,j\rangle} \bigl[ (1 + K_1\sigma_i \sigma_j)\cos(\theta_i - \theta_j) \nonumber \\
 &\hphantom{{}=-\sum_{\langle i,j\rangle} \bigl[} + K_2(\sigma_i - \sigma_j)\sin(\theta_i - \theta_j)\bigr],
\end{align}
where
\begin{align}
 K_1 \equiv& \frac{\sum_{\alpha>1} a_\alpha\bigl[1-\cos(2\phi_{\alpha})\bigr]}{2 + \sum_{\alpha>1} a_\alpha \bigl[1+\cos(2\phi_{\alpha})\bigr]}\\
 K_2 \equiv& \frac{\sum_{\alpha>1} a_\alpha \sin(2\phi_{\alpha})}{2 + \sum_{\alpha>1} a_\alpha \bigl[1+\cos(2\phi_{\alpha})\bigr]}.
\end{align}
This establishes \cref{eq:H_reduced,eq:K1,eq:K2}.

\section{The domain of $(K_1, K_2)$ \label{app:K1K2_domain}}

{The relevant} domain of $(K_1, K_2)$ is given by the area limited by the ellipse
\begin{equation}
 \left[\frac{2}{n-1}K_1 - 1\right]^2 + \left[\frac{2\sqrt{n}}{n-1}K_2\right]^2 \leq 1.
 \label{eq:domain}
\end{equation}
To prove this, it suffices to demonstrate that the equality in \cref{eq:domain} is fulfilled for maximum values of $K_1$ and $K_2$, since smaller values of the left hand side are easily obtained by tuning $a_{\alpha}$.

We note that the maxima of $K_1$ and $K_2$ are obtained when $a_{\alpha} = 1, \forall \alpha$, due to the constant factor $2$ in the denominators of \cref{eq:K1,eq:K2}. Furthermore, since $a_{\alpha} = 1, \forall \alpha$, the maxima are obtained for $\cos(2\phi_{\alpha}) = \cos(2\phi_{\alpha'})$ and $\sin(2\phi_{\alpha}) = \sin(2\phi_{\alpha'})$ $\forall \alpha,\alpha'$. If this were not the case, some bands would contribute more than others, which cannot be the case when the terms are independent and the Hamiltonian is symmetric with respect to band label swapping. We consider the case $\phi_{\alpha} = \phi_{\alpha'} = \phi$, and note that there are equivalent configurations with the $\phi_{\alpha}$'s differing by a sign and/or a factor of $\cpi$.

These considerations simplify the set of maximal $(K_1,K_2)$'s to the one-parameter set
\begin{equation}
  \begin{aligned}
  K_1^\text{max} \equiv& \frac{(n-1)\bigl[1-\cos(2\phi)\bigr]}{2 + (n-1) \bigl[1+\cos(2\phi)\bigr]} \\
  K_2^\text{max} \equiv& \frac{(n-1) \sin(2\phi)}{2 + (n-1) \bigl[1+\cos(2\phi)\bigr]}
  \end{aligned} 
  \label{eq:K_max}
\end{equation}

Inserting \cref{eq:K_max} in the left hand side of \cref{eq:domain} yields 
\begin{align}
 \MoveEqLeft \frac{\bigl[(n-1) + (n+1)\cos(2\phi)\bigr]^2 + 4n \sin^2(2\phi)}{\bigl[(n+1) + (n-1)\cos(2\phi)\bigr]^2} \nonumber\\
 &= \frac{\bigl[(n-1) + (n+1)\cos(2\phi)\bigr]^2 + 4n(1- \cos^2(2\phi))}{\bigl[(n+1) + (n-1)\cos(2\phi)\bigr]^2} \nonumber\\
 &= \frac{(n+1)^2 + 2(n-1)(n+1)\cos(2\phi) + (n-1)^2\cos^2(2\phi)}{\bigl[(n+1) + (n-1)\cos(2\phi)\bigr]^2} \nonumber\\
 &= 1
 \label{eq:maxproof}
\end{align}
This holds for all values of $\phi$. 

From \cref{eq:K_max} we see that $K_1^\text{max}$ takes all values in $[0,n-1]$. Together with \cref{eq:maxproof} this shows that \cref{eq:K_max} indeed is a parametrization of the bounding ellipse of \cref{eq:domain}.

\section{Criterion for a staggered flux phase \label{app:staggered_criterion}}
Provided the parameter $K_2$ is sufficiently large, the reduced model may in principle feature a staggered flux-phase ground state corresponding to an \enquote{antiferomagnetic} (AFM) ordering in the $\groupZ$ sector. Here, we derive the criterion for having such an AFM-ordering ground state. 
Denoting the contribution to the Hamiltonian from a link by $H_{ij}$, this happens when $H_{ij}(\sigma_i\sigma_j = -1) < H_{ij}(\sigma_i\sigma_j = 1)$, or, by \cref{eq:H_reduced},
\begin{equation}
\max_{\Delta\theta}[(1-K_1)\cos \Delta\theta + 2K_2\sin \Delta\theta] > 1 + K_1 ,
\end{equation}
or equivalently
\begin{equation}
\sqrt{ (1-K_1)^2 + (2 K_2)^2 } > 1 + K_1,
\end{equation}
which amounts to 
\begin{equation}
 {K_2}^2 > K_1.
 \label{eq:AFM}
\end{equation}

We now investigate if this criterion can be fulfilled for some region of the domain given by \cref{eq:K1K2_domain}. \Cref{eq:AFM} may be written $K_2^2 = c K_1$, where $c>1$. Inserting this into \cref{eq:K1K2_domain} gives
\begin{equation}
 \left(\frac{2}{n-1}K_1 - 1\right)^2 + \left(\frac{2}{n-1}\right)^2 n cK_1 \leq 1,
\end{equation}
equivalently
\begin{equation}
 K_1^2 + [n(c-1) + 1]K_1 \leq 0
\end{equation}
If $K_1>0$, this can never be fulfilled with $c>1$. Hence, we conclude that the ground state of \cref{eq:H_reduced} (and thus also \cref{eq:H_reduced_gauge}) is one where one has "ferromagnetic" ordering both in the $\groupU$- and $\groupZ$-sectors, when \cref{eq:H_reduced} is viewed as an effective model of an $n$-component strongly frustrated London-superconductor.

\section{Symmetries of the reduced model \cref{eq:H_reduced}  \label{app:symmetries}}

One obvious symmetry of \cref{eq:H_reduced} is that it is invariant under sign-change of $K_2$. This follows immediately from the fact that the operation $K_2 \to - K_2$ can be compensated by letting $\sigma_i \to -\sigma_i$, which is immaterial due to the "up-down" symmetry of the $\groupZ$-sector of the theory. Hence, in the elliptic domain given by \cref{eq:domain}, it suffices to consider $K_2 \geq 0$.  

A less obvious symmetry pertains to $K_1 \in [0,n-1]$. If we reparametrize $K_1$ as follows
\begin{equation}
K_1 \equiv \frac{1-x}{1+x}
\label{eq:re-param}
\end{equation} 
with $x \in [2/n-1,1]$, then the Hamiltonian is symmetric under $x \to -x$. Thus, it suffices to consider $x \in [0,1]$, i.e. $K_1 \in [0,1]$. This is shown as follows. Inserting \cref{eq:re-param} into \cref{eq:H_reduced}, and pulling out a factor $1/(1+x)$ as well as including the factor of inverse temperature $\beta$ appearing in the Boltzmann factor in the canonical partition function, we find
\begin{widetext}
\begin{align}
-\beta H_{ij} & = \frac{\beta}{1+x} \bigl[ (1+x + (1-x) \sigma_i \sigma_j) \cos(\theta_i-\theta_j) + K_2 (1+x) (\sigma_i - \sigma_j) \sin(\theta_i-\theta_j) \bigr]\nonumber \\
        & = \frac{\beta}{1+x} \bigl[ (1 +  \sigma_i \sigma_j ) \cos(\theta_i-\theta_j) + x(1-\sigma_i \sigma_j)\cos(\theta_i-\theta_j)  + K_2 (1+x) (\sigma_i - \sigma_j) \sin(\theta_i-\theta_j) \bigr]\nonumber \\
        & = \frac{2 \beta}{1+x} \bigl[ \delta_{\sigma_i,\sigma_j} \cos(\theta_i-\theta_j) + x \delta_{\sigma_i,-\sigma_j} \cos(\theta_i-\theta_j)  + K_2 (1+x)  \sigma_i\delta_{\sigma_i,-\sigma_j} \sin(\theta_i-\theta_j) \bigr].
\label{eq:H-reparam-1}
\end{align}
Now let $x \to -x$, and at the same time let $\beta/(1-x) \to \beta^{'}/(1+x)$, and $\beta K_2 \to \beta^{'} K_2^{'}$. We then find 
\begin{equation}
-\beta H_{ij} = \frac{2 \beta^{'}}{1+x} \left[ \delta_{\sigma_i,\sigma_j} \cos(\theta_i-\theta_j) 
- x \delta_{\sigma_i,-\sigma_j} \cos(\theta_i-\theta_j)  + K_2^{'} (1+x) \sigma_i \delta_{\sigma_i,-\sigma_j} \sin(\theta_i-\theta_j) \right].
\label{eq:H-reparam-2}
\end{equation}
\end{widetext}
Note that \cref{eq:H-reparam-2} has the same form as \cref{eq:H-reparam-1}, except for the sign-change in front of the $x \delta_{\sigma_i,-\sigma_j} \cos(\theta_i-\theta_j)$-term. {This sign, however, becomes immaterial since for configurations with a given value of $\delta_{\sigma_i,-\sigma_j} \sin(\theta_i-\theta_j)$, there will be two values of $\delta_{\sigma_i,-\sigma_j} \cos(\theta_i-\theta_j)$, and which have opposite signs. Hence, the partition function is invariant under $x \to - x$, up to a rescaling of $\beta$ and $K_2$.}

All in all, it therefore suffices to consider the parameter regime where $(K_1,K_2)$ lies in a part of one quadrant of the domain-ellipse defined by \cref{eq:K1K2_domain}, 
namely $K_1 \in[0,1]$, $K_2 \in [0,(n-1)/2\sqrt{n}]$.

\section{Microcanonical thermodynamics \label{app:microcanonical}}
In this work, we have used Wang--Landau (WL) sampling~\cite{PhysRevLett.86.2050,PhysRevE.64.056101} to investigate the models given by \cref{eq:H_gauge,eq:H_basic,eq:H_reduced_gauge,eq:H_reduced}. The main motivation for this is that broad histogram methods, like the WL algorithm, compares favorably to ordinary canonical sampling in dealing with models having rough energy landscapes (caused by frustration in this case) and (possible) first order phase transitions. Furthermore, the broad range of energies traversed in one WL simulation means that the properties of the model may be determined in a single run, as opposed to a canonical simulation where, if the temperatures of interest are not known a priori, separate computations for a range of temperatures must be performed. 

The WL sampling gives a direct estimate of the density of states $g(E)$ (up to a multiplicative constant) for the Hamiltonian of interest. Based on this, the canonical partition function may be constructed as $Z(\beta) = \sum_E g(E)\exp(-\beta E)$, and an ordinary canonical analysis can be undertaken from then on. In this work, however, we find it more natural, and indeed convenient, to use the \emph{microcanonical} ensemble directly, since $g(E)$ may be viewed as the microcanonical partition function: $g(E) = \sum_\Omega \delta(H(\omega) - E) \equiv \exp S(E)$. Here $\omega \in \Omega$ denotes a field configuration in the set of all possible configurations. $S(E)$ is the \emph{microcanonical} entropy (we use $k_{\text{B}} = 1$ in this work). When normalized with the volume system, $s \equiv S/V$, it is the primary mathematical object of our investigations. 

Although we expect the canonical and microcanonical formalism to give the same results in the thermodynamic limit~\footnote{This is not necessarily true. The canonical approach actually breaks down for first order phase transitions in the thermodynamic limit, a fact which is frequently overseen or forgotten~\cite{gross} }, their finite size scalings (FSS) are different. In general, we expect microcanonical results to be less affected by finite size effects than their canonical counterparts, as the \enquote{finite size smearing} caused by the Boltzmann factor $\exp(-\beta E)$ in the partition function is avoided.\footnote{This \enquote{smearing} is particularly severe close to phase transitions, where states from a broad range of energies give significant contributions to the (finite size) canonical partition function.} Thus, it is often easier to extract the behavior of the system in the thermodynamic limit from small system sizes if a microcanonical approach is used.

In the microcanonical formalism, the inverse temperature $\beta = 1/T$ is \emph{defined} by
\begin{equation}
 \beta(E) \equiv \partial_{E}S(E) = \partial_e s(e),
 \label{eq:beta_def}
\end{equation}
where $e = E/V$. This definition corresponds to the canonical inverse temperature in the thermodynamic limit.\cite{} Here, we use \cref{eq:beta_def} as a definition 
of the \enquote{inverse temperature} even for finite system sizes. Following Eq.~\eqref{eq:beta_def}, the specific heat may be expressed as 
\begin{equation}
 c(e) \equiv \partial_T e = -\beta^2 \partial_\beta e = -\frac{\bigl(\partial_e s(e)\bigr)^2}{\partial_e^2 s(e)}.
 \label{eq:C_def}
\end{equation}

Inspection of $s(e)$ and $\partial_e s(e)$ curves of a given model is used to locate its phase transitions. A second order phase transition can often  be 
identified by a peak in the heat capacity at criticality. This peak corresponds to a small value of the curvature $\partial_e^2 s$, as seen from Eq.~\eqref{eq:C_def}. 
It may either converge to finite height, if the critical exponent $\alpha<0$, or diverge, if $\alpha>0$, as $L\to \infty$. Hence, a second order phase transition 
manifests itself as a \enquote{kink} at a critical value $e=e_{c}$ in a plot of $\partial_e s(e)$, and the \enquote{flatness} of the kink tells us whether the 
transition is of type $\alpha>0$ ($\partial_e^2 s \to 0$) or $\alpha < 0$ ($\partial_e^2 s \to \text{constant} > 0$).

In the models we consider in this work, we expect a 3D $\groupZ$ (Ising) universality class transition and a 3D $\groupU$ (XY) universality class phase transition to take place. 
For the 3D $\groupZ$ universality class $\alpha = \num{0.1096(5)} > 0$~\cite{PhysRevE.65.066127}, so we expect $\partial_e s(e_\text{c})$ to become flat in the thermodynamic 
limit. For the 3D $\groupU$ universality class $\alpha = \num{-0.0146(8)} < 0$~\cite{PhysRevB.63.214503} and hence the slope of $\partial_e s(e_\text{c})$ will always be 
finite in this case. Based on this, we expect $\groupU$ transitions to be harder to accurately determine than $\groupZ$ transitions in the $\partial_e s$-plots. 
See \cref{fig:dsde_example} for an example.

\begin{figure}[tbp]
\includegraphics{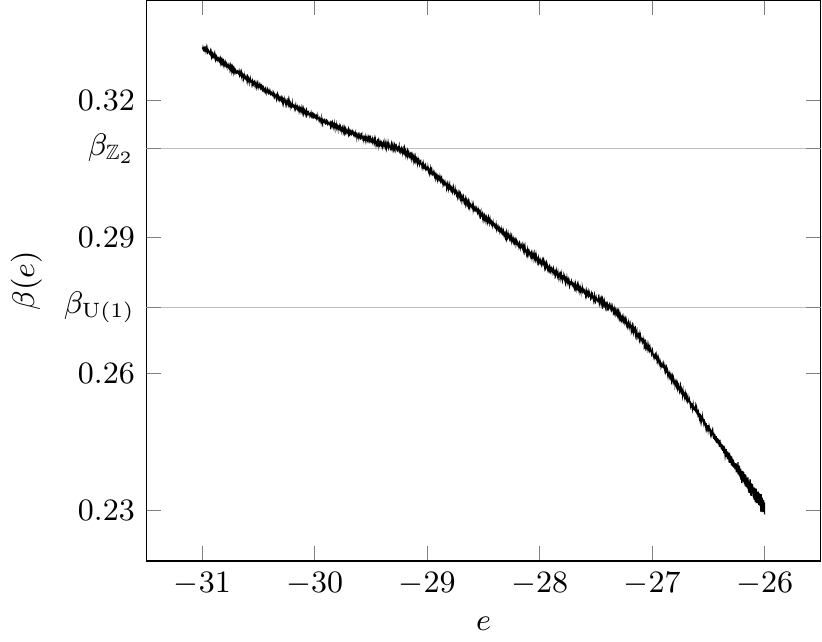}
\caption{Example of a $\beta(e)$ curve obtained from nummerical differentiation of $s(e)$. The critical $\beta(e)$ values are associated with the kinks in the curve, the 
$\groupZ$ kink being closer to horizontal than the $\groupU$ kink. The data are from a $L=40$ simulation of the full 
model without gauge field, \cref{eq:H_basic}, with $g_{12} = g_{13} = 20$ and $g_{23} = 17.5$. }
\label{fig:dsde_example}
\end{figure}

It is primarily in investigating first order phase transitions the microcanonical approach shows itself far superior to the canonical one. A first order transition is easily identified as a convex intruder (where $\partial_e^2 s > 0$) in the otherwise concave $s(e)$~\footnote{As long as one takes the possibility of some exotic finite size effects into account; see e.g. Ref. \onlinecite{PhysRevE.74.011108}.}. The energy range where $s(e)$ is convex determines the latent heat, $\Delta e$, as shown in \cref{fig:s_beta_shift_example}. Note that field configurations within this range are exponentially suppressed in canonical simulations, which is why a canonical formulation may be ill-suited for investigating first order phase transitions. \cite{}

\begin{figure}[tbp]
\includegraphics{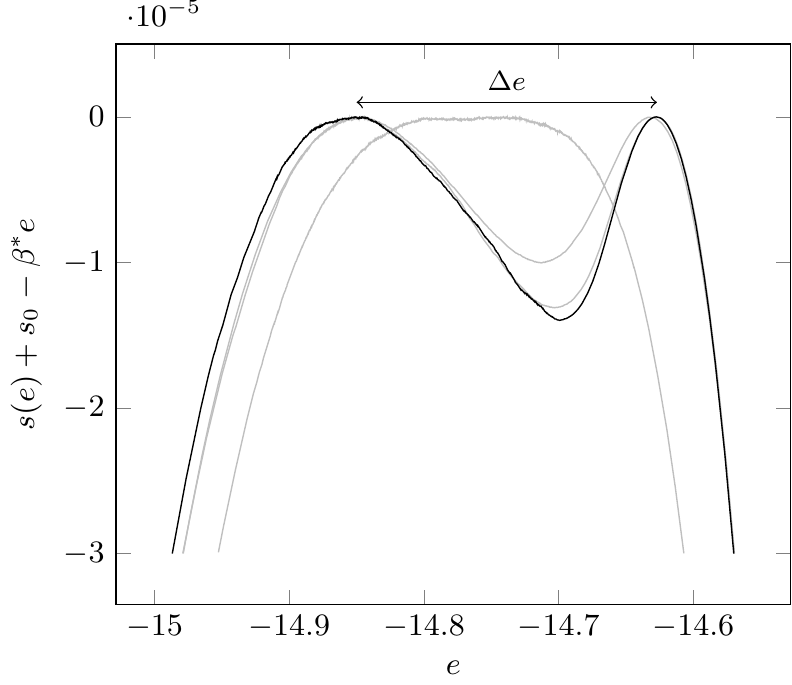}
\caption{A plot of the convex intruder in the entropy $s(e)$, here shifted by a factor $s_0 - \beta^*e$ to make it visible, of the full model without gauge field, \cref{eq:H_basic}, with $g_{12} = g_{13} = g_{23} = 10$. The latent heat $\Delta e$ is indicated. A range of system sizes are shown, displaying the finite size effects involved. $s_0$ is an arbitrary (unknown) constant, $\beta^*$ is the inverse \enquote{finite size transition temperature} (which makes the two peaks equal in height). The black curve is from a $L = 80$ simulation. Here $\beta^* = 0.3168395$. The gray curves are from $L = \set{60,40,20}$ simulations, with $\beta^* = \set{0.3168256, 0.3167752, 0.3165}$. Note that the $L=20$ system size is too small to reveal the true nature of the phase transition.}
\label{fig:s_beta_shift_example}
\end{figure}

\section{Numerical techniques \label{numerical_techniques}}
A shared memory Wang--Landau (WL) algorithm~\cite{Zhan2008} with a combined minimum histogram-flatness criterion and a $1/t$ change of the update factor~\cite{PhysRevE.75.046701,2007JChPh.127r4105B} was used to sample the approximate density of states. Since the Hamiltonians contain continuous degrees of freedom, the true densities of states are continuous as well. The density of states was therefore approximated by a densely binned discrete set. We do not expect this to affect the results in any significant way. Only a small interval of the entire energy domain is of interest, so the WL walkers were restricted to a \enquote{window} containing this interval (and extending a little outside this to avoid boundary effects and improve ergodicity). The method of Ref. \onlinecite{PhysRevE.67.067102} was used in order to minimize window boundary effects. 32 WL walkers, each with its own field copy, sampled the window simultaneously with data race allowed. Pseudorandom numbers were generated by the Mersenne--Twister algorithm~\cite{Matsumoto_1998_ACM_TMCS}.

$\partial_e s$ was obtained by numerical differentiation. Since numerical differentiation is an ill-posed problem, any method of differentiation must be a tradeoff between noise suppression and the possibility of introducing systematic errors. A simple finite difference approximation, like $\partial_e s(e_i) \approx (s(e_{i+1}) - s(e_i))/(e_{i+1} - e_i)$ proved too noise sensitive for our purposes. Instead, a second order differentiator kernel of P. Holoborodko~\cite{Holoborodko}, of width $2\times50 + 1 = 101$, turned out to be acceptable, and was used. The width was deemed negligible compared to the total number of bins ($\mathcal{O}(10^5)$ -- $\mathcal{O}(10^6)$) and the (assumed) smooth structure of the true $s(e)$ curve. Hence, we may regard the result as a good approximation to the true derivative at any given point. 

The final data were obtained manually by inspection of either $\partial_e s$ or $s(e) - \beta^* e$ curves. $\beta^*$ is a manually tuned parameter. The reason for shifting the $s(e)$ curve in this way is to facilitate the  identification of the convex intruder of a first order transition. This visual approach was deemed to be more convenient and reliable than an algorithmic, possibly noise sensitive, method.

\section{Additional Numerical Techniques and Checks \label{app:new_numerical}}
In order to check the validity of the microcanonical WL approach described in \cref{numerical_techniques}, a \emph{canonical} ensemble, local update Monte Carlo scheme was implemented for the model of \cref{eq:H_basic}. The \enquote{Fast Linear Algorithm} (FLA) of Ref. \onlinecite{Loison_et_al_2004} was used. It proved to be a significant improvement over traditional Metropolis-Hastings sampling, and is to our knowledge the best canonical algorithm available for this problem.\footnote{The frustration in the system prevented us from using a nonlocal cluster algorithm.} This approach does not match the performance of the WL algorithm when dealing with first order phase transitions, but is expected to be competitive for second order phase transitions and in the limit of weak Josephson coupling. It also has the advantage that it can be parallelized on a grid. 

To probe the phase transitions we used the order parameter \enquote{magnetizations}
\begin{equation}
m_{\groupZ}   \equiv  N\inv \sum_i \sigma_i, 
\end{equation}
and 
\begin{align}
 m_{\groupU} &\equiv \tfrac{1}{7}\bigl[ m_{\groupU,\theta_1} + m_{\groupU,\theta_2} + m_{\groupU,\theta_3} \nonumber\\
 &\hphantom{\equiv\bigl[{}}+ m_{\groupU,\theta_1 + \theta_2} + m_{\groupU,\theta_2 + \theta_3} + m_{\groupU,\theta_3 + \theta_1} \nonumber\\
 &\hphantom{\equiv\bigl[{}}+ m_{\groupU,\theta_1 + \theta_2 + \theta_3} \bigr],
\end{align} 
where
\begin{equation}
 m_{\groupU,x} \equiv N\inv \bigl| \sum_j \exp(\i x_j) \bigr|.
\end{equation}

We use the Binder ratio~\cite{Sandvik_2010},
\begin{equation}
 R \equiv \frac{\avg{m^4}}{\avg{m^2}^2},
\end{equation}
to detect phase transitions. The Binder ratio displays a nonanalytical jump at the phase transition in the thermodynamical limit, and has the useful property of being only mildly affected by finite size effects.

We performed a simulation using $g_{12} = g_{13} = 20$ and $g_{32}=19$, with the result shown in \cref{fig:sanity_check}. This is in good agreement with \cref{fig:full_nogauge_g23}.

\begin{figure}[tbp]
\includegraphics{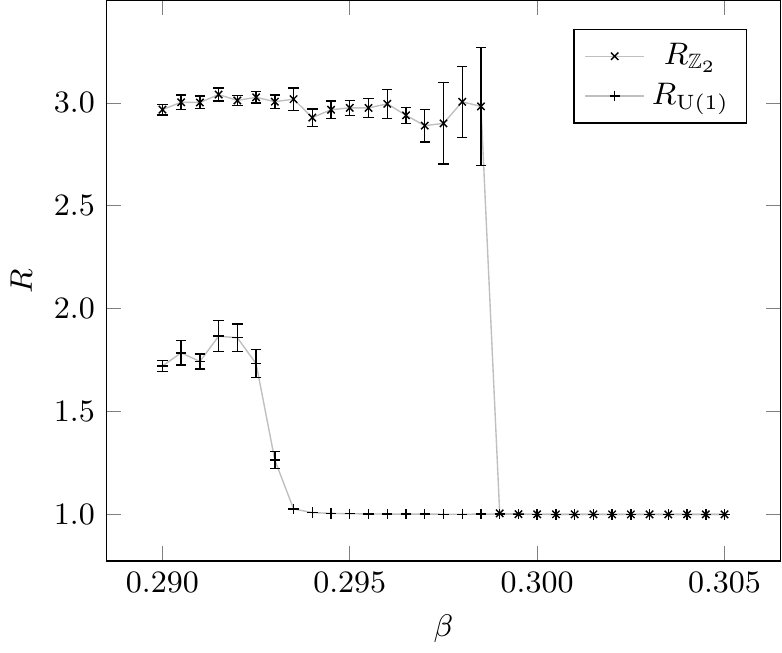}
\caption{Binder ratios from a canonical simulation of the full model without gauge field, \cref{eq:H_basic}. $g_{12} = g_{13} = 20$, $g_{32}=19$ and $L=128$. The critical couplings $\beta_{\groupZ} \approx 0.299$ and $\beta_{\groupU} \approx 0.293$ are in good agreement with \cref{fig:full_nogauge_g23}.}
\label{fig:sanity_check}
\end{figure}

{Our main focus in this paper concerns the regime of Josephson-couplings  $\{ g \} > 1$. In addition, we have performed some computations in the low $g$-regime to see if we could detect a splitting of the $\groupUZ$ transition line shown in \cref{fig:full_nogauge_g}a to two separate $\groupU$ and $\groupZ$ transition lines. At the lowest values which we simulated $g_{12} = g_{13} = g_{23} = g = 0.001$, we have seen no sign of any such splitting for  the system sizes which are accessible to us, see see \cref{fig:split_search}. As discussed in the text, in the low-$g$ limit, the growth of the domain wall width should require large lattice sizes to detect $\groupZ$ transition. Thus, although we have not detected two separate transitions for these small values of $g$ that we considered at the largest system sizes we have been able to simulate, we have not ruled out that splitting occurs in the thermodynamic limit.}
\begin{figure}[tbp]
\includegraphics{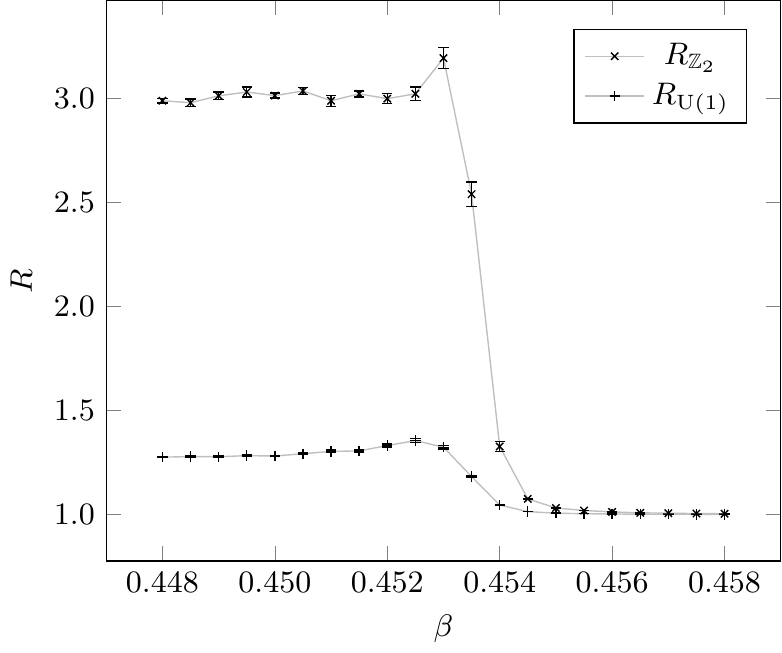}
\caption{Binder ratios from a canonical simulation of the full model without gauge field, \cref{eq:H_basic}. $g_{12} = g_{13} = g_{32}= 0.001$ and $L=128$. Based on this simulation it is not possible to conclude that the transition couplings $\beta_{\groupZ}$ and $\beta_{\groupU}$ are different.}
\label{fig:split_search}
\end{figure}

 \bibliography{references}

\begin{thebibliography}{38}%
\makeatletter
\providecommand \@ifxundefined [1]{%
 \@ifx{#1\undefined}
}%
\providecommand \@ifnum [1]{%
 \ifnum #1\expandafter \@firstoftwo
 \else \expandafter \@secondoftwo
 \fi
}%
\providecommand \@ifx [1]{%
 \ifx #1\expandafter \@firstoftwo
 \else \expandafter \@secondoftwo
 \fi
}%
\providecommand \natexlab [1]{#1}%
\providecommand \enquote  [1]{``#1''}%
\providecommand \bibnamefont  [1]{#1}%
\providecommand \bibfnamefont [1]{#1}%
\providecommand \citenamefont [1]{#1}%
\providecommand \href@noop [0]{\@secondoftwo}%
\providecommand \href [0]{\begingroup \@sanitize@url \@href}%
\providecommand \@href[1]{\@@startlink{#1}\@@href}%
\providecommand \@@href[1]{\endgroup#1\@@endlink}%
\providecommand \@sanitize@url [0]{\catcode `\\12\catcode `\$12\catcode
  `\&12\catcode `\#12\catcode `\^12\catcode `\_12\catcode `\%12\relax}%
\providecommand \@@startlink[1]{}%
\providecommand \@@endlink[0]{}%
\providecommand \url  [0]{\begingroup\@sanitize@url \@url }%
\providecommand \@url [1]{\endgroup\@href {#1}{\urlprefix }}%
\providecommand \urlprefix  [0]{URL }%
\providecommand \Eprint [0]{\href }%
\providecommand \doibase [0]{http://dx.doi.org/}%
\providecommand \selectlanguage [0]{\@gobble}%
\providecommand \bibinfo  [0]{\@secondoftwo}%
\providecommand \bibfield  [0]{\@secondoftwo}%
\providecommand \translation [1]{[#1]}%
\providecommand \BibitemOpen [0]{}%
\providecommand \bibitemStop [0]{}%
\providecommand \bibitemNoStop [0]{.\EOS\space}%
\providecommand \EOS [0]{\spacefactor3000\relax}%
\providecommand \BibitemShut  [1]{\csname bibitem#1\endcsname}%
\let\auto@bib@innerbib\@empty
\bibitem [{\citenamefont {Kamihara}\ \emph {et~al.}(2008)\citenamefont
  {Kamihara}, \citenamefont {Watanabe}, \citenamefont {Hirano},\ and\
  \citenamefont {Hosono}}]{iron}%
  \BibitemOpen
  \bibfield  {author} {\bibinfo {author} {\bibfnamefont {Y.}~\bibnamefont
  {Kamihara}}, \bibinfo {author} {\bibfnamefont {T.}~\bibnamefont {Watanabe}},
  \bibinfo {author} {\bibfnamefont {M.}~\bibnamefont {Hirano}}, \ and\ \bibinfo
  {author} {\bibfnamefont {H.}~\bibnamefont {Hosono}},\ }\href {\doibase
  10.1021/ja800073m} {\bibfield  {journal} {\bibinfo  {journal} {J. Am. Chem.
  Soc.}\ }\textbf {\bibinfo {volume} {130}},\ \bibinfo {pages} {3296} (\bibinfo
  {year} {2008})}\BibitemShut {NoStop}%
\bibitem [{\citenamefont {Ng}\ and\ \citenamefont {Nagaosa}(2009)}]{nagaosa}%
  \BibitemOpen
  \bibfield  {author} {\bibinfo {author} {\bibfnamefont {T.~K.}\ \bibnamefont
  {Ng}}\ and\ \bibinfo {author} {\bibfnamefont {N.}~\bibnamefont {Nagaosa}},\
  }\href {\doibase 10.1209/0295-5075/87/17003} {\bibfield  {journal} {\bibinfo
  {journal} {Europhys. Lett.}\ }\textbf {\bibinfo {volume} {87}},\ \bibinfo
  {pages} {17003} (\bibinfo {year} {2009})}\BibitemShut {NoStop}%
\bibitem [{\citenamefont {Stanev}\ and\ \citenamefont
  {Te\vaccent{s}anovi\'{c}}(2010)}]{zlatko}%
  \BibitemOpen
  \bibfield  {author} {\bibinfo {author} {\bibfnamefont {V.}~\bibnamefont
  {Stanev}}\ and\ \bibinfo {author} {\bibfnamefont {Z.}~\bibnamefont
  {Te\vaccent{s}anovi\'{c}}},\ }\href {\doibase 10.1103/PhysRevB.81.134522}
  {\bibfield  {journal} {\bibinfo  {journal} {Phys. Rev. B}\ }\textbf {\bibinfo
  {volume} {81}},\ \bibinfo {pages} {134522} (\bibinfo {year}
  {2010})}\BibitemShut {NoStop}%
\bibitem [{\citenamefont {Carlstr\"om}\ \emph {et~al.}(2011)\citenamefont
  {Carlstr\"om}, \citenamefont {Garaud},\ and\ \citenamefont
  {Babaev}}]{johan3}%
  \BibitemOpen
  \bibfield  {author} {\bibinfo {author} {\bibfnamefont {J.}~\bibnamefont
  {Carlstr\"om}}, \bibinfo {author} {\bibfnamefont {J.}~\bibnamefont {Garaud}},
  \ and\ \bibinfo {author} {\bibfnamefont {E.}~\bibnamefont {Babaev}},\ }\href
  {\doibase 10.1103/PhysRevB.84.134518} {\bibfield  {journal} {\bibinfo
  {journal} {Phys. Rev. B}\ }\textbf {\bibinfo {volume} {84}},\ \bibinfo
  {pages} {134518} (\bibinfo {year} {2011})}\BibitemShut {NoStop}%
\bibitem [{\citenamefont {{Maiti}}\ and\ \citenamefont
  {{Chubukov}}(2013)}]{maiti}%
  \BibitemOpen
  \bibfield  {author} {\bibinfo {author} {\bibfnamefont {S.}~\bibnamefont
  {{Maiti}}}\ and\ \bibinfo {author} {\bibfnamefont {A.~V.}\ \bibnamefont
  {{Chubukov}}},\ }\href {\doibase 10.1103/PhysRevB.87.144511} {\bibfield
  {journal} {\bibinfo  {journal} {\prb}\ }\textbf {\bibinfo {volume} {87}},\
  \bibinfo {eid} {144511} (\bibinfo {year} {2013})},\ \Eprint
  {http://arxiv.org/abs/1302.2964} {arXiv:1302.2964 [cond-mat.supr-con]}
  \BibitemShut {NoStop}%
\bibitem [{\citenamefont {Mukherjee}\ and\ \citenamefont
  {Agterberg}(2011)}]{agterberg2011}%
  \BibitemOpen
  \bibfield  {author} {\bibinfo {author} {\bibfnamefont {S.}~\bibnamefont
  {Mukherjee}}\ and\ \bibinfo {author} {\bibfnamefont {D.~F.}\ \bibnamefont
  {Agterberg}},\ }\href {\doibase 10.1103/PhysRevB.84.134520} {\bibfield
  {journal} {\bibinfo  {journal} {Phys. Rev. B}\ }\textbf {\bibinfo {volume}
  {84}},\ \bibinfo {pages} {134520} (\bibinfo {year} {2011})}\BibitemShut
  {NoStop}%
\bibitem [{\citenamefont {{Lee}}\ \emph {et~al.}(2009)\citenamefont {{Lee}},
  \citenamefont {{Zhang}},\ and\ \citenamefont {{Wu}}}]{other1}%
  \BibitemOpen
  \bibfield  {author} {\bibinfo {author} {\bibfnamefont {W.-C.}\ \bibnamefont
  {{Lee}}}, \bibinfo {author} {\bibfnamefont {S.-C.}\ \bibnamefont {{Zhang}}},
  \ and\ \bibinfo {author} {\bibfnamefont {C.}~\bibnamefont {{Wu}}},\ }\href
  {\doibase 10.1103/PhysRevLett.102.217002} {\bibfield  {journal} {\bibinfo
  {journal} {Physical Review Letters}\ }\textbf {\bibinfo {volume} {102}},\
  \bibinfo {eid} {217002} (\bibinfo {year} {2009})},\ \Eprint
  {http://arxiv.org/abs/0810.0887} {arXiv:0810.0887 [cond-mat.supr-con]}
  \BibitemShut {NoStop}%
\bibitem [{\citenamefont {{Platt}}\ \emph {et~al.}(2012)\citenamefont
  {{Platt}}, \citenamefont {{Thomale}}, \citenamefont {{Honerkamp}},
  \citenamefont {{Zhang}},\ and\ \citenamefont {{Hanke}}}]{other2}%
  \BibitemOpen
  \bibfield  {author} {\bibinfo {author} {\bibfnamefont {C.}~\bibnamefont
  {{Platt}}}, \bibinfo {author} {\bibfnamefont {R.}~\bibnamefont {{Thomale}}},
  \bibinfo {author} {\bibfnamefont {C.}~\bibnamefont {{Honerkamp}}}, \bibinfo
  {author} {\bibfnamefont {S.-C.}\ \bibnamefont {{Zhang}}}, \ and\ \bibinfo
  {author} {\bibfnamefont {W.}~\bibnamefont {{Hanke}}},\ }\href {\doibase
  10.1103/PhysRevB.85.180502} {\bibfield  {journal} {\bibinfo  {journal}
  {\prb}\ }\textbf {\bibinfo {volume} {85}},\ \bibinfo {eid} {180502} (\bibinfo
  {year} {2012})},\ \Eprint {http://arxiv.org/abs/1106.5964} {arXiv:1106.5964
  [cond-mat.supr-con]} \BibitemShut {NoStop}%
\bibitem [{\citenamefont {{Lin}}\ and\ \citenamefont {{Hu}}(2012)}]{lin}%
  \BibitemOpen
  \bibfield  {author} {\bibinfo {author} {\bibfnamefont {S.-Z.}\ \bibnamefont
  {{Lin}}}\ and\ \bibinfo {author} {\bibfnamefont {X.}~\bibnamefont {{Hu}}},\
  }\href {\doibase 10.1103/PhysRevLett.108.177005} {\bibfield  {journal}
  {\bibinfo  {journal} {Physical Review Letters}\ }\textbf {\bibinfo {volume}
  {108}},\ \bibinfo {eid} {177005} (\bibinfo {year} {2012})},\ \Eprint
  {http://arxiv.org/abs/1107.0814} {arXiv:1107.0814 [cond-mat.supr-con]}
  \BibitemShut {NoStop}%
\bibitem [{\citenamefont {{Stanev}}(2012)}]{stanev}%
  \BibitemOpen
  \bibfield  {author} {\bibinfo {author} {\bibfnamefont {V.}~\bibnamefont
  {{Stanev}}},\ }\href {\doibase 10.1103/PhysRevB.85.174520} {\bibfield
  {journal} {\bibinfo  {journal} {\prb}\ }\textbf {\bibinfo {volume} {85}},\
  \bibinfo {eid} {174520} (\bibinfo {year} {2012})},\ \Eprint
  {http://arxiv.org/abs/1108.2501} {arXiv:1108.2501 [cond-mat.supr-con]}
  \BibitemShut {NoStop}%
\bibitem [{\citenamefont {{Marciani}}\ \emph {et~al.}(2013)\citenamefont
  {{Marciani}}, \citenamefont {{Fanfarillo}}, \citenamefont {{Castellani}},\
  and\ \citenamefont {{Benfatto}}}]{marciani}%
  \BibitemOpen
  \bibfield  {author} {\bibinfo {author} {\bibfnamefont {M.}~\bibnamefont
  {{Marciani}}}, \bibinfo {author} {\bibfnamefont {L.}~\bibnamefont
  {{Fanfarillo}}}, \bibinfo {author} {\bibfnamefont {C.}~\bibnamefont
  {{Castellani}}}, \ and\ \bibinfo {author} {\bibfnamefont {L.}~\bibnamefont
  {{Benfatto}}},\ }\href@noop {} {\bibfield  {journal} {\bibinfo  {journal}
  {ArXiv e-prints}\ } (\bibinfo {year} {2013})},\ \Eprint
  {http://arxiv.org/abs/1306.5545} {arXiv:1306.5545 [cond-mat.supr-con]}
  \BibitemShut {NoStop}%
\bibitem [{\citenamefont {{Leggett}}(1966)}]{leggett}%
  \BibitemOpen
  \bibfield  {author} {\bibinfo {author} {\bibfnamefont {A.~J.}\ \bibnamefont
  {{Leggett}}},\ }\href {\doibase 10.1143/PTP.36.901} {\bibfield  {journal}
  {\bibinfo  {journal} {Progress of Theoretical Physics}\ }\textbf {\bibinfo
  {volume} {36}},\ \bibinfo {pages} {901} (\bibinfo {year} {1966})}\BibitemShut
  {NoStop}%
\bibitem [{\citenamefont {{Garaud}}\ \emph {et~al.}(2011)\citenamefont
  {{Garaud}}, \citenamefont {{Carlstr{\"o}m}},\ and\ \citenamefont
  {{Babaev}}}]{cp21}%
  \BibitemOpen
  \bibfield  {author} {\bibinfo {author} {\bibfnamefont {J.}~\bibnamefont
  {{Garaud}}}, \bibinfo {author} {\bibfnamefont {J.}~\bibnamefont
  {{Carlstr{\"o}m}}}, \ and\ \bibinfo {author} {\bibfnamefont {E.}~\bibnamefont
  {{Babaev}}},\ }\href {\doibase 10.1103/PhysRevLett.107.197001} {\bibfield
  {journal} {\bibinfo  {journal} {Physical Review Letters}\ }\textbf {\bibinfo
  {volume} {107}},\ \bibinfo {eid} {197001} (\bibinfo {year} {2011})},\ \Eprint
  {http://arxiv.org/abs/1107.0995} {arXiv:1107.0995 [cond-mat.supr-con]}
  \BibitemShut {NoStop}%
\bibitem [{\citenamefont {{Garaud}}\ \emph {et~al.}(2013)\citenamefont
  {{Garaud}}, \citenamefont {{Carlstr{\"o}m}}, \citenamefont {{Babaev}},\ and\
  \citenamefont {{Speight}}}]{cp22}%
  \BibitemOpen
  \bibfield  {author} {\bibinfo {author} {\bibfnamefont {J.}~\bibnamefont
  {{Garaud}}}, \bibinfo {author} {\bibfnamefont {J.}~\bibnamefont
  {{Carlstr{\"o}m}}}, \bibinfo {author} {\bibfnamefont {E.}~\bibnamefont
  {{Babaev}}}, \ and\ \bibinfo {author} {\bibfnamefont {M.}~\bibnamefont
  {{Speight}}},\ }\href {\doibase 10.1103/PhysRevB.87.014507} {\bibfield
  {journal} {\bibinfo  {journal} {\prb}\ }\textbf {\bibinfo {volume} {87}},\
  \bibinfo {eid} {014507} (\bibinfo {year} {2013})},\ \Eprint
  {http://arxiv.org/abs/1211.4342} {arXiv:1211.4342 [cond-mat.supr-con]}
  \BibitemShut {NoStop}%
\bibitem [{\citenamefont {{Garaud}}\ and\ \citenamefont
  {{Babaev}}(2013)}]{cp33}%
  \BibitemOpen
  \bibfield  {author} {\bibinfo {author} {\bibfnamefont {J.}~\bibnamefont
  {{Garaud}}}\ and\ \bibinfo {author} {\bibfnamefont {E.}~\bibnamefont
  {{Babaev}}},\ }\href@noop {} {\bibfield  {journal} {\bibinfo  {journal}
  {ArXiv e-prints}\ } (\bibinfo {year} {2013})},\ \Eprint
  {http://arxiv.org/abs/1308.3220} {arXiv:1308.3220 [cond-mat.supr-con]}
  \BibitemShut {NoStop}%
\bibitem [{\citenamefont {{Silaev}}\ and\ \citenamefont
  {{Babaev}}(2013)}]{silaevv}%
  \BibitemOpen
  \bibfield  {author} {\bibinfo {author} {\bibfnamefont {M.}~\bibnamefont
  {{Silaev}}}\ and\ \bibinfo {author} {\bibfnamefont {E.}~\bibnamefont
  {{Babaev}}},\ }\href@noop {} {\bibfield  {journal} {\bibinfo  {journal}
  {ArXiv e-prints}\ } (\bibinfo {year} {2013})},\ \Eprint
  {http://arxiv.org/abs/1306.6159} {arXiv:1306.6159 [cond-mat.supr-con]}
  \BibitemShut {NoStop}%
\bibitem [{\citenamefont {{Bojesen}}\ \emph {et~al.}(2013)\citenamefont
  {{Bojesen}}, \citenamefont {{Babaev}},\ and\ \citenamefont
  {{Sudb{\o}}}}]{2d}%
  \BibitemOpen
  \bibfield  {author} {\bibinfo {author} {\bibfnamefont {T.}~\bibnamefont
  {{Bojesen}}}, \bibinfo {author} {\bibfnamefont {E.}~\bibnamefont {{Babaev}}},
  \ and\ \bibinfo {author} {\bibfnamefont {A.}~\bibnamefont {{Sudb{\o}}}},\
  }\href@noop {} {\bibfield  {journal} {\bibinfo  {journal} {ArXiv e-prints}\ }
  (\bibinfo {year} {2013})},\ \Eprint {http://arxiv.org/abs/1306.2313}
  {arXiv:1306.2313 [cond-mat.supr-con]} \BibitemShut {NoStop}%
\bibitem [{\citenamefont {Weston}\ and\ \citenamefont
  {Babaev}(2013)}]{Weston_2013}%
  \BibitemOpen
  \bibfield  {author} {\bibinfo {author} {\bibfnamefont {D.}~\bibnamefont
  {Weston}}\ and\ \bibinfo {author} {\bibfnamefont {E.}~\bibnamefont
  {Babaev}},\ }\href@noop {} {\bibfield  {journal} {\bibinfo  {journal}
  {Unknown Journal}\ } (\bibinfo {year} {2013})},\ \Eprint
  {http://arxiv.org/abs/1306.3179} {arxiv:1306.3179} \BibitemShut {NoStop}%
\bibitem [{\citenamefont {{Lan}}\ \emph {et~al.}(2012)\citenamefont {{Lan}},
  \citenamefont {{Hsieh}},\ and\ \citenamefont {{Kao}}}]{2012arXiv1211.0780L}%
  \BibitemOpen
  \bibfield  {author} {\bibinfo {author} {\bibfnamefont {T.-Y.}\ \bibnamefont
  {{Lan}}}, \bibinfo {author} {\bibfnamefont {Y.-D.}\ \bibnamefont {{Hsieh}}},
  \ and\ \bibinfo {author} {\bibfnamefont {Y.-J.}\ \bibnamefont {{Kao}}},\
  }\href@noop {} {\bibfield  {journal} {\bibinfo  {journal} {ArXiv e-prints}\ }
  (\bibinfo {year} {2012})},\ \Eprint {http://arxiv.org/abs/1211.0780}
  {arXiv:1211.0780 [cond-mat.stat-mech]} \BibitemShut {NoStop}%
\bibitem [{Note1()}]{Note1}%
  \BibitemOpen
  \bibinfo {note} {The ground state contribution of the Josephson term, i.e.
  the minimum of $\DOTSB \sum@ \slimits@ _{\alpha >\alpha '} g_{\alpha \alpha
  '}\protect \qopname \relax o{cos}(\theta _\alpha - \theta _{\alpha '})$, is
  reached when $\theta _2 = \theta _3$ when $g_{23} \leq g/2$, when $g_{12} =
  g_{13} = g$.}\BibitemShut {Stop}%
\bibitem [{\citenamefont {Wang}\ and\ \citenamefont
  {Landau}(2001{\natexlab{a}})}]{PhysRevLett.86.2050}%
  \BibitemOpen
  \bibfield  {author} {\bibinfo {author} {\bibfnamefont {F.}~\bibnamefont
  {Wang}}\ and\ \bibinfo {author} {\bibfnamefont {D.~P.}\ \bibnamefont
  {Landau}},\ }\href {\doibase 10.1103/PhysRevLett.86.2050} {\bibfield
  {journal} {\bibinfo  {journal} {Phys. Rev. Lett.}\ }\textbf {\bibinfo
  {volume} {86}},\ \bibinfo {pages} {2050} (\bibinfo {year}
  {2001}{\natexlab{a}})}\BibitemShut {NoStop}%
\bibitem [{\citenamefont {Wang}\ and\ \citenamefont
  {Landau}(2001{\natexlab{b}})}]{PhysRevE.64.056101}%
  \BibitemOpen
  \bibfield  {author} {\bibinfo {author} {\bibfnamefont {F.}~\bibnamefont
  {Wang}}\ and\ \bibinfo {author} {\bibfnamefont {D.~P.}\ \bibnamefont
  {Landau}},\ }\href {\doibase 10.1103/PhysRevE.64.056101} {\bibfield
  {journal} {\bibinfo  {journal} {Phys. Rev. E}\ }\textbf {\bibinfo {volume}
  {64}},\ \bibinfo {pages} {056101} (\bibinfo {year}
  {2001}{\natexlab{b}})}\BibitemShut {NoStop}%
\bibitem [{Note2()}]{Note2}%
  \BibitemOpen
  \bibinfo {note} {This is not necessarily true. The canonical approach
  actually breaks down for first order phase transitions in the thermodynamic
  limit, a fact which is frequently overseen or forgotten~\cite
  {gross}}\BibitemShut {NoStop}%
\bibitem [{Note3()}]{Note3}%
  \BibitemOpen
  \bibinfo {note} {This \enquote {smearing} is particularly severe close to
  phase transitions, where states from a broad range of energies give
  significant contributions to the (finite size) canonical partition
  function.}\BibitemShut {Stop}%
\bibitem [{\citenamefont {Campostrini}\ \emph {et~al.}(2002)\citenamefont
  {Campostrini}, \citenamefont {Pelissetto}, \citenamefont {Rossi},\ and\
  \citenamefont {Vicari}}]{PhysRevE.65.066127}%
  \BibitemOpen
  \bibfield  {author} {\bibinfo {author} {\bibfnamefont {M.}~\bibnamefont
  {Campostrini}}, \bibinfo {author} {\bibfnamefont {A.}~\bibnamefont
  {Pelissetto}}, \bibinfo {author} {\bibfnamefont {P.}~\bibnamefont {Rossi}}, \
  and\ \bibinfo {author} {\bibfnamefont {E.}~\bibnamefont {Vicari}},\ }\href
  {\doibase 10.1103/PhysRevE.65.066127} {\bibfield  {journal} {\bibinfo
  {journal} {Phys. Rev. E}\ }\textbf {\bibinfo {volume} {65}},\ \bibinfo
  {pages} {066127} (\bibinfo {year} {2002})}\BibitemShut {NoStop}%
\bibitem [{\citenamefont {Campostrini}\ \emph {et~al.}(2001)\citenamefont
  {Campostrini}, \citenamefont {Hasenbusch}, \citenamefont {Pelissetto},
  \citenamefont {Rossi},\ and\ \citenamefont {Vicari}}]{PhysRevB.63.214503}%
  \BibitemOpen
  \bibfield  {author} {\bibinfo {author} {\bibfnamefont {M.}~\bibnamefont
  {Campostrini}}, \bibinfo {author} {\bibfnamefont {M.}~\bibnamefont
  {Hasenbusch}}, \bibinfo {author} {\bibfnamefont {A.}~\bibnamefont
  {Pelissetto}}, \bibinfo {author} {\bibfnamefont {P.}~\bibnamefont {Rossi}}, \
  and\ \bibinfo {author} {\bibfnamefont {E.}~\bibnamefont {Vicari}},\ }\href
  {\doibase 10.1103/PhysRevB.63.214503} {\bibfield  {journal} {\bibinfo
  {journal} {Phys. Rev. B}\ }\textbf {\bibinfo {volume} {63}},\ \bibinfo
  {pages} {214503} (\bibinfo {year} {2001})}\BibitemShut {NoStop}%
\bibitem [{Note4()}]{Note4}%
  \BibitemOpen
  \bibinfo {note} {As long as one takes the possibility of some exotic finite
  size effects into account; see e.g. Ref. \protect \rev@citealpnum
  {PhysRevE.74.011108}.}\BibitemShut {Stop}%
\bibitem [{\citenamefont {Zhan}(2008)}]{Zhan2008}%
  \BibitemOpen
  \bibfield  {author} {\bibinfo {author} {\bibfnamefont {L.}~\bibnamefont
  {Zhan}},\ }\href {\doibase http://dx.doi.org/10.1016/j.cpc.2008.04.002}
  {\bibfield  {journal} {\bibinfo  {journal} {Computer Physics Communications}\
  }\textbf {\bibinfo {volume} {179}},\ \bibinfo {pages} {339 } (\bibinfo {year}
  {2008})}\BibitemShut {NoStop}%
\bibitem [{\citenamefont {Belardinelli}\ and\ \citenamefont
  {Pereyra}(2007)}]{PhysRevE.75.046701}%
  \BibitemOpen
  \bibfield  {author} {\bibinfo {author} {\bibfnamefont {R.~E.}\ \bibnamefont
  {Belardinelli}}\ and\ \bibinfo {author} {\bibfnamefont {V.~D.}\ \bibnamefont
  {Pereyra}},\ }\href {\doibase 10.1103/PhysRevE.75.046701} {\bibfield
  {journal} {\bibinfo  {journal} {Phys. Rev. E}\ }\textbf {\bibinfo {volume}
  {75}},\ \bibinfo {pages} {046701} (\bibinfo {year} {2007})}\BibitemShut
  {NoStop}%
\bibitem [{\citenamefont {{Belardinelli}}\ and\ \citenamefont
  {{Pereyra}}(2007)}]{2007JChPh.127r4105B}%
  \BibitemOpen
  \bibfield  {author} {\bibinfo {author} {\bibfnamefont {R.~E.}\ \bibnamefont
  {{Belardinelli}}}\ and\ \bibinfo {author} {\bibfnamefont {V.~D.}\
  \bibnamefont {{Pereyra}}},\ }\href {\doibase 10.1063/1.2803061} {\bibfield
  {journal} {\bibinfo  {journal} {\jcp}\ }\textbf {\bibinfo {volume} {127}},\
  \bibinfo {pages} {184105} (\bibinfo {year} {2007})},\ \Eprint
  {http://arxiv.org/abs/arXiv:cond-mat/0702414} {arXiv:cond-mat/0702414}
  \BibitemShut {NoStop}%
\bibitem [{\citenamefont {Schulz}\ \emph {et~al.}(2003)\citenamefont {Schulz},
  \citenamefont {Binder}, \citenamefont {M\"uller},\ and\ \citenamefont
  {Landau}}]{PhysRevE.67.067102}%
  \BibitemOpen
  \bibfield  {author} {\bibinfo {author} {\bibfnamefont {B.~J.}\ \bibnamefont
  {Schulz}}, \bibinfo {author} {\bibfnamefont {K.}~\bibnamefont {Binder}},
  \bibinfo {author} {\bibfnamefont {M.}~\bibnamefont {M\"uller}}, \ and\
  \bibinfo {author} {\bibfnamefont {D.~P.}\ \bibnamefont {Landau}},\ }\href
  {\doibase 10.1103/PhysRevE.67.067102} {\bibfield  {journal} {\bibinfo
  {journal} {Phys. Rev. E}\ }\textbf {\bibinfo {volume} {67}},\ \bibinfo
  {pages} {067102} (\bibinfo {year} {2003})}\BibitemShut {NoStop}%
\bibitem [{\citenamefont {Matsumoto}\ and\ \citenamefont
  {Nishimura}(1998)}]{Matsumoto_1998_ACM_TMCS}%
  \BibitemOpen
  \bibfield  {author} {\bibinfo {author} {\bibfnamefont {M.}~\bibnamefont
  {Matsumoto}}\ and\ \bibinfo {author} {\bibfnamefont {T.}~\bibnamefont
  {Nishimura}},\ }\href {\doibase 10.1145/272991.272995} {\bibfield  {journal}
  {\bibinfo  {journal} {ACM Trans. Model. Comput. Simul.}\ }\textbf {\bibinfo
  {volume} {8}},\ \bibinfo {pages} {3} (\bibinfo {year} {1998})}\BibitemShut
  {NoStop}%
\bibitem [{\citenamefont {Holoborodko}(2008)}]{Holoborodko}%
  \BibitemOpen
  \bibfield  {author} {\bibinfo {author} {\bibfnamefont {P.}~\bibnamefont
  {Holoborodko}},\ }\href@noop {} {\enquote {\bibinfo {title} {Smooth noise
  robust differentiators},}\ }\bibinfo {howpublished}
  {http://www.holoborodko.com/pavel/numerical-methods/numerical-derivative/smooth-low-noise-differentiators/}
  (\bibinfo {year} {2008})\BibitemShut {NoStop}%
\bibitem [{\citenamefont {Loison}\ \emph {et~al.}(2004)\citenamefont {Loison},
  \citenamefont {Qin}, \citenamefont {Schotte},\ and\ \citenamefont
  {Jin}}]{Loison_et_al_2004}%
  \BibitemOpen
  \bibfield  {author} {\bibinfo {author} {\bibfnamefont {D.}~\bibnamefont
  {Loison}}, \bibinfo {author} {\bibfnamefont {C.}~\bibnamefont {Qin}},
  \bibinfo {author} {\bibfnamefont {K.}~\bibnamefont {Schotte}}, \ and\
  \bibinfo {author} {\bibfnamefont {X.}~\bibnamefont {Jin}},\ }\href {\doibase
  10.1140/epjb/e2004-00332-5} {\bibfield  {journal} {\bibinfo  {journal} {The
  European Physical Journal B - Condensed Matter and Complex Systems}\ }\textbf
  {\bibinfo {volume} {41}},\ \bibinfo {pages} {395} (\bibinfo {year}
  {2004})}\BibitemShut {NoStop}%
\bibitem [{Note5()}]{Note5}%
  \BibitemOpen
  \bibinfo {note} {The frustration in the system prevented us from using a
  nonlocal cluster algorithm.}\BibitemShut {Stop}%
\bibitem [{\citenamefont {{Sandvik}}(2010)}]{Sandvik_2010}%
  \BibitemOpen
  \bibfield  {author} {\bibinfo {author} {\bibfnamefont {A.~W.}\ \bibnamefont
  {{Sandvik}}},\ }in\ \href {\doibase 10.1063/1.3518900} {\emph {\bibinfo
  {booktitle} {American Institute of Physics Conference Series}}},\ \bibinfo
  {series} {American Institute of Physics Conference Series}, Vol.\ \bibinfo
  {volume} {1297},\ \bibinfo {editor} {edited by\ \bibinfo {editor}
  {\bibfnamefont {A.}~\bibnamefont {{Avella}}}\ and\ \bibinfo {editor}
  {\bibfnamefont {F.}~\bibnamefont {{Mancini}}}}\ (\bibinfo {year} {2010})\
  pp.\ \bibinfo {pages} {135--338},\ \Eprint {http://arxiv.org/abs/1101.3281}
  {arXiv:1101.3281 [cond-mat.str-el]} \BibitemShut {NoStop}%
\bibitem [{\citenamefont {Gross}(2001)}]{gross}%
  \BibitemOpen
  \bibfield  {author} {\bibinfo {author} {\bibfnamefont {D.~H.~E.}\
  \bibnamefont {Gross}},\ }\href@noop {} {\emph {\bibinfo {title}
  {{Microcanonical Thermodynamics -- Phase Transitions in \enquote{Small}
  Systems}}}},\ \bibinfo {series} {World Scientific Lecture Notes in Physics},
  Vol.~\bibinfo {volume} {66}\ (\bibinfo  {publisher} {World Scientific},\
  \bibinfo {year} {2001})\BibitemShut {NoStop}%
\bibitem [{\citenamefont {Behringer}\ and\ \citenamefont
  {Pleimling}(2006)}]{PhysRevE.74.011108}%
  \BibitemOpen
  \bibfield  {author} {\bibinfo {author} {\bibfnamefont {H.}~\bibnamefont
  {Behringer}}\ and\ \bibinfo {author} {\bibfnamefont {M.}~\bibnamefont
  {Pleimling}},\ }\href {\doibase 10.1103/PhysRevE.74.011108} {\bibfield
  {journal} {\bibinfo  {journal} {Phys. Rev. E}\ }\textbf {\bibinfo {volume}
  {74}},\ \bibinfo {pages} {011108} (\bibinfo {year} {2006})}\BibitemShut
  {NoStop}%
\end{thebibliography}%

\end{document}